\newif\ifsubmode
\submodefalse

\newif\ifprintfig
\printfigtrue

\ifsubmode
  \documentstyle[12pt,aasms4,epsf]{article}
  \received{}
  \accepted{}
  \journalid{}{}
  \articleid{}{}
\else
  \documentstyle[11pt,aaspp4,epsf]{article}
  \slugcomment{{\it Astronomical Journal, submitted May 2001.}}
\fi

\lefthead{van der Marel}
\righthead{Magellanic Cloud Structure II}

\newcommand{\etal}{{et al.~}}

\newcommand{\lta}{\lesssim}
\newcommand{\gta}{\gtrsim}

\newcommand{\kms}{\>{\rm km}\,{\rm s}^{-1}}
\newcommand{\pc}{\>{\rm pc}}
\newcommand{\kpc}{\>{\rm kpc}}

\newcommand{\Msun}{\>{\rm M_{\odot}}}

\newcommand{\chour}{^{\rm h}\>}
\newcommand{\cmin}{^{\rm m}\>}

\begin{document}

\title{Magellanic Cloud Structure from Near-IR Surveys II:\\ 
Star Count Maps and the Intrinsic Elongation of the LMC}

\author{Roeland P.~van der Marel}
\affil{Space Telescope Science Institute, 3700 San Martin Drive, 
       Baltimore, MD 21218}



\ifsubmode\else
\clearpage\fi


\ifsubmode\else
\baselineskip=14pt
\fi


\begin{abstract}
I construct a near-IR star count map of the LMC and demonstrate, using
the viewing angles derived in Paper~I, that the LMC is intrinsically
elongated. I argue that this is due to the tidal force from the Milky
Way.

The near-IR data from the 2MASS and DENIS surveys are ideally suited
for studies of LMC structure, because of the large statistics and
insensitivity to dust absorption. The survey data are used to create a
star count map of RGB and AGB stars. The resulting LMC image shows the
well-known bar, but is otherwise quite smooth. Ellipse fitting is used
for quantitative analysis. The radial number density profile is
approximately exponential with a scale-length $r_d
\approx 1.3$--$1.5 \kpc$. However, there is an excess density at large 
radii that may be due to the tidal effect of the Milky Way. The
position angle and ellipticity profile both show large radial
variations, but converge to ${\rm PA}_{\rm maj} = 189.3^{\circ}
\pm 1.4^{\circ}$ and $\epsilon = 0.199 \pm 0.008$ for $r \gta 5^{\circ}$.
At large radii the image is influenced by viewing perspective (i.e.,
one side of the inclined LMC plane being closer to us than the other).
This causes a drift of the center of the star count contours towards
the near side of the plane. The observed drift is consistent with the
position angle $\Theta = 122.5^{\circ} \pm 8.3^{\circ}$ of the line of
nodes inferred in Paper~I.

The fact that $\Theta$ differs from ${\rm PA}_{\rm maj}$ indicates
that the LMC disk is not circular. Deprojection shows that the LMC has
an intrinsic ellipticity $\epsilon'' = 0.31$ in its outer parts,
considerably larger than typical for disk galaxies. The outer contours
have a more-or-less common center, which lies $\sim 0.4 \kpc$ from the
center of the bar. Neither agrees with the kinematic center of the HI
gas disk. The LMC is elongated in the general direction of the
Galactic center, and is elongated perpendicular to the Magellanic
Stream and the velocity vector of the LMC center of mass. This
suggests that the elongation of the LMC has been induced by the tidal
force of the Milky Way.

The position angle of the line of nodes differs from the position
angle $\Theta_{\rm max}$ of the line of maximum line of sight velocity
gradient. Results from HI gas and discrete tracers indicate that
$\Theta_{\rm max} - \Theta = 20^{\circ}$--$60^{\circ}$. This could be
due to one or more of the following: (a) streaming along non-circular
orbits in the elongated disk; (b) uncertainties in the transverse
motion of the LMC center of mass, which can translate into a spurious
solid-body rotation component in the observed velocity field; (c) an
additional solid body rotation component in the observed velocity
field due to precession and nutation of the LMC disk as it orbits the
Milky Way, which is expected on theoretical grounds.
\end{abstract}



\clearpage



\section{Introduction}
\label{s:intro}

The structure of the Magellanic Clouds has been a topic of intense
study for many years (as reviewed by Westerlund 1997). This subject is
of fundamental importance to several questions, including the study of
the Galactic dark halo through modeling of the Magellanic Stream and
the tidal disruption of the Magellanic Clouds (e.g., Moore
\& Davis 1994; Lin, Jones \& Klemola 1995; Gardiner \& Noguchi 1996)
and the study of compact objects in the dark halo through microlensing
studies (e.g., Alcock \etal 2000; Lasserre \etal 2000; Udalski \etal
1999). The structure of the Large Magellanic Cloud (LMC) in particular
has been studied by many authors, using almost every possible tracer,
including: optically detected starlight (de Vaucouleurs \& Freeman
1973; Bothun \& Thompson 1988; Schmidt-Kaler \& Gochermann 1992);
stellar clusters (Lynga \& Westerlund 1963; Freeman \etal 1983; Kontizas \etal
1990; Schommer \etal 1992); HII regions, supergiants, or planetary nebulae
(Feitzinger, Isserstedt \& Schmidt-Kaler 1977; Meatheringham \etal
1988); carbon stars (Kunkel \etal 1997; Graff \etal 2000; Alves \&
Nelson 2000); red giants (Zaritsky \& Lin 1997; Zaritsky \etal 1999);
HI emission (McGee \& Milton 1966; Rohlfs \etal 1984; Luks \& Rohlfs
1992; Kim \etal 1998); or non-thermal radio emission (Alvarez, Aparici
\& May 1987).

Recently, two important near-IR surveys have become available for
studies of the Magellanic Clouds, the Two Micron All Sky Survey
(2MASS; e.g., Skrutskie 1998) and the Deep Near-Infrared Southern Sky
Survey (DENIS; e.g., Epchtein \etal 1997). These surveys are perfect
for a study of LMC structure. Near-IR data is quite insensitive to
dust absorption, which is a major complicating factor in optical
studies (Zaritsky, Harris \& Thompson 1997; Zaritsky 1999). The
surveys have superb statistics with of the order of a million
stars. Also, the observational strategy with three near-IR bands ($J$,
$H$ and $K_s$ in the 2MASS survey; $I$, $J$ and $K_s$ in the DENIS
survey) allows clear separation of different stellar populations. In
particular, the data are ideal for studies of evolved Red Giant Branch
(RGB) and Asymptotic Giant Branch (AGB) stars, which emit much of
their light in the near-IR. This is important for studies of LMC
structure, because these intermediate-age and old stars are more
likely to trace the underlying mass distribution of the LMC disk than
younger populations that dominate the light in optical images (e.g.,
de Vaucouleurs
\& Freeman 1973). The present paper is the second in a series of
studies of the Magellanic Clouds using the near-IR 2MASS and DENIS
surveys. The series focuses on the structure of the Clouds, and
addresses topics that are either different or more in-depth than
previous work with the same data sets (Cioni, Habing \& Israel 2000b;
Cioni \etal 2000c; Nikolaev \& Weinberg 2000; Weinberg
\& Nikolaev 2000).

The generally accepted consensus on the structure of the LMC is that
it is an approximately planar galaxy that is circular at large radii.
At smaller radii there is an off-center bar. The planarity of the LMC
is supported by a large body of evidence, as reviewed in Section~9.3
of Paper~I (van der Marel \& Cioni 2001). It has been a topic of
debate whether the LMC contains secondary populations that do not
reside in the main disk plane (e.g., Luks \& Rohlfs 1992; Zaritsky \&
Lin 1997; Zaritsky \etal 1999; Weinberg \& Nikolaev 2000; Zhao \&
Evans 2000; this issue will be further addressed in Paper~III, van der
Marel 2001), but the planar geometry of the primary LMC population is
not generally called into question. On the other hand, the consensus
that the LMC disk is circular (at large radii) is based entirely on
assumptions, and has never been tested. The projected shape and
kinematics of the LMC are determined both by the intrinsic shape of
the LMC, and the angles under which it is viewed (the inclination
angle $i$, and the position angle $\Theta$ of the line of
nodes). Almost all existing studies have assumed that the LMC disk is
circular, to obtain estimates of the viewing angles (see the review of
this subject in Paper~I).

However, there exists an alternative, and much more accurate method to
estimate the LMC viewing angles. It does not make any assumptions
about the intrinsic shape of the LMC, but merely uses simple geometry.
Because the LMC plane is inclined with respect to the plane of the
sky, one side of it is closer to us than the other. This causes
sinusoidal variations in the apparent brightness of LMC tracers as a
function of position angle on the sky. In Paper~I we sought such
variations in the apparent magnitude of features in the near-IR
color-magnitude diagrams (CMDs) extracted from the DENIS and 2MASS
surveys. The expected sinusoidal variations are confidently detected
with a peak-to-peak amplitude of $\sim 0.25$ mag; they are seen both
for AGB stars (using the mode of their luminosity function) and for
RGB stars (using the tip of their luminosity function), and are seen
consistently in all three of the photometric bands in both surveys.
The inferred viewing angles are $i = 34.7^{\circ} \pm 6.2^{\circ}$ and
$\Theta = 122.5^{\circ} \pm 8.3^{\circ}$ (defined such that the near
side of the LMC is at $\Theta_{\rm near} \equiv \Theta -
90^{\circ}$). The inclination angle is consistent with previous
estimates. However, the line-of-nodes position angle is not.
Photometric and kinematic studies have generally found that the major
axis position angle $\Theta_{\rm maj}$ of the spatial distribution of
tracers on the sky, and also the position angle $\Theta_{\rm max}$ of
the line of maximum gradient in the LMC velocity field, fall somewhere
in the range $140^{\circ}$--$190^{\circ}$. For a circular disk one
would expect that $\Theta = \Theta_{\rm maj} = \Theta_{\rm
max}$. Consequently, the LMC disk cannot be circular, but must be
elongated.

The present paper addresses the nature and origin of the intrinsic
elongation of the LMC disk. Section~\ref{s:maps} presents a map of the
stellar number density of RGB and AGB stars on the projected plane of
the sky. Section~\ref{s:isoshape} presents a quantitative analysis of
this map using ellipse fitting to the number density
contours. Section~\ref{s:faceon} presents a deprojected face-on map of
the LMC, obtained with the viewing angles from Paper~I; this map too
is analyzed using ellipse fitting. Section~\ref{s:perspective}
describes the influence of viewing perspective on the appearance of
the LMC on the projected plane of the sky. Section~\ref{s:elong}
discusses the inferred intrinsic elongation of the LMC, and the
evidence that it might have been tidally
induced. Section~\ref{s:kinematics} address the observed kinematics of
the LMC in the context of our new knowledge of LMC
structure. Section~\ref{s:conc} summarizes the main findings.

\section{Sky Maps of the Stellar Number Density}
\label{s:maps}

\subsection{Coordinates and Map Creation} 
\label{ss:coords}

I use the near-IR stellar catalogs compiled by the 2MASS and DENIS
surveys to create maps of the stellar number density on the sky.
There are many ways to project the surface of the celestial sphere
onto a plane (e.g., Calabretta 1992). The present analysis uses the
so-called `zenithal equidistant projection'. The projection origin
${\cal O}$ with celestial coordinates $(\alpha_0,\delta_0)$ is chosen
to be at RA = $5\chour 29\cmin$ and DEC = $-69.5^{\circ}$ (here and
henceforth all coordinates are J2000.0). This corresponds roughly to
the center of the outer number density contours of the LMC; however,
the exact choice of the projection origin is fairly arbitrary and does
not affect any of the subsequent analysis at a significant
level. Angular coordinates $(\rho,\phi)$ are defined on the celestial
sphere, where $\rho$ is the angular distance between a point
$(\alpha,\delta)$ and the origin $(\alpha_0,\delta_0)$, and $\phi$ is
the position angle of the point $(\alpha,\delta)$ with respect to the
origin; by convention, $\phi$ is measured counterclockwise starting
from the axis that runs in the direction of decreasing RA at constant
declination $\delta_0$.  The coordinates $(\rho,\phi)$ can be
calculated for any $(\alpha,\delta)$ using the formulae of spherical
trigonometry (see Paper~I). Cartesian projection coordinates $(X,Y)$
are defined according to
\begin{equation}
  X(\alpha,\delta) \equiv \rho \cos \phi , \qquad
  Y(\alpha,\delta) \equiv \rho \sin \phi .
\label{projdef}
\end{equation}
The positive $X$-axis points West, and the positive $Y$-axis points
North, in agreement with the usual astronomical convention.

For each star in a given list the projection coordinates $(X,Y)$ can
be calculated from the known celestial coordinates
$(\alpha,\delta)$. The stars are then binned into square pixels of
size $\Delta X = \Delta Y \equiv \Delta = 0.05^{\circ}$. The solid
angle $\Omega$ that a pixel subtends on the sky is the integral over
${\rm d}\Omega \equiv \sin \rho \> {\rm d}\rho \> {\rm d}\phi$, which
can be approximated as $\Omega = (\sin \rho / \rho) \> \Delta X \>
\Delta Y$. The number of stars in a pixel is divided by its solid
angle to obtain $\Sigma$, the observed number density of stars per
square degree. The resulting maps are smoothed using a Gaussian with a
dispersion of 1 pixel ($\sigma = 0.05^{\circ}$) to reduce noise.

\subsection{Correction for Distance Bias}
\label{ss:distcor}

Stellar surveys are generally complete only to a certain limiting
magnitude. This causes a distance bias in number density maps,
especially for large nearby galaxies such as the LMC. The plane of the
LMC is inclined with respect to the plane of the sky, and stars on the
near side of the LMC plane appear brighter than those on the far side
(see Paper~I for a detailed analysis of this effect). Consequently, a
catalog that is limited to a fixed apparent magnitude contains
intrinsically fainter stars on the near side of the LMC than on the
far side. This artificially boosts the observed number density on the
near side.

The LMC viewing angles determined in Paper~I can be used to correct
for this bias. Let a star be observed at position $(\alpha,\delta)$
with apparent magnitude $m$. If it is assumed that the star resides in
the plane of the LMC, then the formulae of Paper~I allow one to
calculate the ratio $D/D_0$ of the distance $D$ to the star and the
distance $D_0$ to the LMC center. One can then also calculate the
`distance-corrected apparent magnitude' $m_0 = m - 5\log(D/D_0)$. This
is the apparent magnitude that the star would have had at the distance
$D_0$. In the following I correct for distance biases by restricting
the samples to stars for which the distance-corrected apparent
magnitude is brighter than some fixed cutoff magnitude $m_{\rm
lim}$. The cutoff magnitude is chosen such that $m_{\rm lim} +
5\log(D/D_0)$ is brighter than the survey magnitude limit for the
entire LMC area (within which $| 5\log(D/D_0) | \lta 0.3$ mag,
cf.~Figure~1 of Paper~1).

While the correction for distance bias in the sample selection
improves the quality of the analysis, the results were not found to be
critically dependent on this correction. Even without application of
the correction all the main results of the present paper still hold.

\subsection{Selection of RGB and AGB stars}
\label{ss:RGBAGBselec}

Both the 2MASS and DENIS survey obtained data in three near-IR
bands. The color-magnitude diagrams that can be constructed from these
data allow one to select stars in particular evolutionary phases.
Weinberg \& Nikolaev (2000) used 2MASS data to show star count maps
for various groups of stars in the $(J-K_s,K_s)$ CMD. Cioni, Habing \&
Israel (2000b) used DENIS data to show star count maps for various
groups of stars in the $(I-J,I)$ CMD. The primary goal of the present
paper is to quantitatively study the structure of the LMC. In this
context one is not so much interested in young stars and the regions
of star formation, but more in the underlying mass distribution. The
latter is best traced by intermediate-age and old stars, and in
particular stars in the RGB and AGB evolutionary phases. The analysis
is therefore restricted to these stars.

Separate LMC number density maps were created from the 2MASS and
DENIS data. For 2MASS I used the Second 2MASS Incremental Data
Release. Sources were extracted from the 2MASS Point Source Catalog
that are in the LMC region of the sky, restricting the extraction to
those stars detected in all three of the 2MASS bands with no special
error flags.  Figure~\ref{f:cmds}a shows the $(J-K_s,K_s)$
CMD. Nikolaev \& Weinberg (2000) presented a detailed discussion of
the general features of this CMD. They divided the CMD in disjunct
areas, and discussed the stellar types that contribute most to each
area. The solid lines in Figure~\ref{f:cmds}a mark the boundary of the
region within which I used stars to create an LMC number density
map. This region encompasses the regions named E, F, G, J, and K by
Nikolaev \& Weinberg (2000), plus small parts of D and L. It consists
almost exclusively of RGB and AGB stars in the LMC, with at the
faintest magnitudes possibly a small contamination by Galactic dwarfs
and background galaxies. A distance-corrected apparent magnitude limit
$K_{s,{\rm lim}} = 14$ was applied.

For DENIS I used the DENIS Catalog towards the Magellanic Clouds
(DCMC; Cioni \etal 2000a), with the improved photometric zeropoint
calibration discussed in Paper~I. Sources were extracted that were
detected in the $I$ and $J$ bands, independent of whether or not they
were detected in $K_s$. Sources with non-optimal values of any of the
DCMC data-quality flags were not excluded, to optimize the statistics
of the sample. Figure~\ref{f:cmds}b shows the $(I-J,J)$ CMD, which was
discussed previously in Cioni \etal (2000a,b,c) and Paper~I. The solid
lines mark the boundary of the region within which I used stars to
create an LMC number density map. This region encompasses the regions
named B and C by Cioni \etal (2000b), in which RGB and AGB stars are
again believed to be the main contributors. A distance-corrected
apparent magnitude limit $I_{\rm lim} = 16$ was applied. RGB stars
with $K_s = 14$ have $I-K_s \approx 2.2$ (see Figure~2 of Paper~I), so
the samples extracted from the 2MASS ($K_{s,{\rm lim}} = 14$) and
DENIS data ($I_{\rm lim} = 16$) have roughly similar depth.

\subsection{Resulting Maps}
\label{ss:maps}

Figure~\ref{f:maps}a shows the number density distribution inferred
from the 2MASS data. The two dark vertical rectangles are regions for
which no data is available in the 2MASS Second Incremental Data
Release. Otherwise the 2MASS map looks very smooth and regular, with
no evidence for obvious artifacts.

Figure~\ref{f:maps}b shows the surface number density distribution
inferred from the DENIS data. There are no missing regions for this
data set, but the map does show some artifacts in the form of features
along lines of constant declination (which run close to vertical). The
LMC area in the DCMC consists of 119 individual scan strips that are
12 arcmin wide in RA and $30^{\circ}$ long in declination (e.g., Cioni
\etal 2000a). The artifacts in the map are the result of imperfections
in the catalog data for some ($\lta 10$\%) of the individual scan
strips.

The Galactic foreground contribution was estimated for each map by
fitting a linear number density distribution $\Sigma_{\rm fore} = a +
bX + cY$ to the data in areas of approximately one square degree in
the four corners of each map. These foreground contribution models
were subtracted.

The foreground-subtracted LMC maps from the 2MASS and DENIS surveys
agree very well, despite the somewhat different CMD selection criteria
for the two surveys. Blinking of the two maps on a computer screen
shows good agreement in both large-scale and small-scale features
(with the exception of the missing regions and artifacts discussed
above). It is therefore justified to combine the two maps into one
higher quality map. To create such a map I started from the 2MASS map,
which appears to have the highest cosmetic quality. The regions with
missing 2MASS data were then filled in using the data in the DENIS
map. The latter were first scaled to the same average surface number
density as the 2MASS data, which is necessary because the CMD
selection criteria were different for the two maps
(cf.~Figure~\ref{f:cmds}). Figure~\ref{f:maps}c shows the combined
map, which forms the basis for the subsequent discussions.

\subsection{Near-IR Morphology of the LMC}
\label{ss:morphology}

The morphology of the LMC has been discussed by many previous authors
(e.g., Westerlund 1997). The most obvious structure is of course the
central bar, which is clearly visible in the near-IR map in
Figure~\ref{f:maps}c. de Vaucouleurs \& Freeman (1973) identified
various morphological features in the LMC from deep optical images,
and listed several irregular, asymmetric and `embryonic' spiral arms,
and an outer loop. Some of these features are also visible in the
near-IR map (especially when viewed in various contrasts on a computer
screen). At the eastern end of the bar there is a feature, termed `B3'
by de Vaucouleurs \& Freeman, which winds counterclockwise by $\sim
60^{\circ}$ around the center. At the western end of the bar there is
a weaker but longer feature termed `B1' that winds counterclockwise by
$\sim 140^{\circ}$. These and other fainter features give the LMC
morphology some hints of spiral structure. However, all of these
features have very low contrast with respect to their surroundings,
and there is certainly no well organized spiral pattern in the LMC. In
fact, the most noticeable feature of the near-IR map in
Figure~\ref{f:maps}c is really the smoothness of the LMC number
density distribution for these intermediate-age and old stars. This
was noticed previously by e.g., Cioni \etal (2000b) and Weinberg \&
Nikolaev (2000), and stands in stark contrast with the much more
irregular and clumpy distribution of, e.g., younger stars and HII
regions in the LMC (Westerlund 1997). This property makes the near-IR
AGB/RGB map well-suited for quantitative analysis.

\section{Ellipse Fit Analysis of the Sky Map} 
\label{s:isoshape}

To analyze the observed number density distribution I used the task
`ellipse' in the IRAF STSDAS package, which fits ellipses to the
isodensity contours. The task is based on the algorithm described in
Jedrzejewski (1987). It assumes that each contour can to lowest order
be well approximated by an ellipse, which was found to be an adequate
assumption (the higher-order Fourier coefficients for the LMC are no
larger than a few percent). The ellipse center was not kept fixed, but
was fit separately for each contour; so it is not assumed that the LMC
is symmetric (which it is not). Figure~\ref{f:ellpar} shows the
parameters of the best-fitting ellipses for the combined 2MASS+DENIS
map shown in Figure~\ref{f:maps}c. The radius $r$ along the abscissa
is the semi-major axis length of an ellipse in the projected
coordinates defined by equation~(\ref{projdef}).

The subsequent discussions and analysis are restricted to the combined
2MASS+DENIS map. However, a similar ellipse fit analysis was also
performed for the 2MASS and DENIS maps separately. Although these
analyses are complicated by the missing regions (2MASS) and artifacts
(DENIS) in these maps, they do yield results that are fully consistent
with each other and with the results from the combined map (especially
for the outer contours, for which small scale irregularities tend to
have little influence). All main the results reported in this paper
are robust, and are seen in both surveys.

\subsection{The Radial Number Density Profile}
\label{ss:radprofile}

Figure~\ref{f:ellpar}a shows the radial number density profile, which
declines monotonically with radius. The dashed line shows the best
fitting exponential profile, $\Sigma = \Sigma_0
\exp(-r/r_d)$, which has scale length $r_d = 1.44^{\circ}$. 
At the distance of the LMC ($\sim 51 \kpc$), 1 degree on the sky
corresponds to $0.89 \kpc$, so that this corresponds to $r_d = 1.28
\kpc$.

While an exponential model can be fit to the data, it does not provide
a very good description. The data show clear deviations from the best
fit model. At intermediate radii, $1^{\circ} \lta r \lta 3.5^{\circ}$,
the brightness profile falls steeper than at both smaller and larger
radii. An exponential model fit to this radial range yields $r_d =
1.12^{\circ}$. At a radius $r \approx 3.5^{\circ}$ the number density
has an upturn, and beyond this radius $\Sigma(r)$ falls slower than at
smaller radii. An exponential model fit to the radial range
$3.5^{\circ} \lta r \lta 6^{\circ}$ yields $r_d = 2.61^{\circ}$. At
radii $r \gta 6^{\circ}$ the number density starts falling more
steeply again. At $r = 8^{\circ}$ the density has fallen to one-third
of the Galactic foreground contribution that was subtracted from the
map (dotted line in Figure~\ref{f:ellpar}a).

\subsection{The Ellipticity and Position Angle Profiles}
\label{ss:ellPA}

Figures~\ref{f:ellpar}c,e show the radial profiles of the ellipticity
$\epsilon$ (defined as $1 - q$, where $q$ is the axial ratio) and the
major axis position angle ${\rm PA}_{\rm maj}$ (measured
counterclockwise from North, defined modulo $180^{\circ}$). Both
profiles show pronounced variations as function of distance from the
LMC center.

The morphology out to $r \approx 3^{\circ}$ is dominated by the bar,
which has a position angle that falls from ${\rm PA}_{\rm maj} \approx
126^{\circ}$ at $r=0.4^{\circ}$ to ${\rm PA}_{\rm maj} \approx
112^{\circ}$ at $r=2.0^{\circ}$, before it starts to rise again. The
ellipticity of the bar first rises from $\epsilon \approx 0.43$ at
$r=0.4^{\circ}$ to $\epsilon \approx 0.67$ at $r=1.0^{\circ}$, and
then falls again to lower values.

The region $2^{\circ} \lta r \lta 4^{\circ}$ marks the smooth
transition between the bar and the outer morphology. In this region
the ellipticity shows a pronounced drop, and ${\rm PA}_{\rm maj}$
twists by $\sim 80^{\circ}$.

At radii $r \gta 4^{\circ}$ the contour shapes converge to an
approximately constant position angle and ellipticity. The RMS
variations in position angle and ellipticity at $r > 5^{\circ}$ are
$2.7^{\circ}$ and $0.016$, respectively. The average values and their
formal errors over this radial range are ${\rm PA}_{\rm maj} =
189.3^{\circ} \pm 1.4^{\circ}$ and $\epsilon = 0.199 \pm 0.008$.

\subsection{The Drift of the Contour Center}
\label{ss:center}

The center of a fitted ellipse is denoted
$(X_0,Y_0)$. Figures~\ref{f:ellpar}b,d show the radial profiles of
$X_0$ and $Y_0$. Figure~\ref{f:ellpar}f shows how $(X_0,Y_0)$ drifts
in the projected plane of the sky as the semi-major axis length of the
contours changes.  

The center $(X_0,Y_0)$ is measured with respect to the origin ${\cal
O}$ of the zenithal equidistant projection defined by
equation~(\ref{projdef}), which was chosen to be at RA = $5\chour
29\cmin$ and DEC = $-69.5^{\circ}$. The maximum surface density,
determined as the average center of the inner few contours, occurs at
$(X_0,Y_0) = (+0.34^{\circ},-0.28^{\circ})$, with an error of
approximately $0.01^{\circ}$ in both coordinates. This corresponds to
RA = $5\chour 25.1\cmin \pm 0.1\cmin$ and DEC = $-69^{\circ} 47' \pm
1'$.

The bar does not have a fixed center, and nor do the outer
contours. Starting from the position of maximum surface density, the
center of the bar first drifts West by $\sim 0.7^{\circ}$ until the
semi-major axis length has increased to $r
\approx 4^{\circ}$, and it then drifts back East by
a similar amount until the semi-major axis length has increased to $r
\approx 5.4^{\circ}$. For the outer contours the center
drifts to the North-North-East. The drift of the outer contours can be
explained as the result of viewing perspective, as discussed in
Section~\ref{s:perspective}. The center of the outermost contour at $r
\approx 8^{\circ}$ is displaced by $\sim 1^{\circ}$ from the center of
the bar.

\subsection{Comparison to previous work}
\label{ss:compprev}

A large number of results on the morphology and structure of the LMC
are available from previous work, using a variety of tracers (see
references in Section~\ref{s:intro}). While every tracer has its pros
and cons, none of these previous studies has had all the important
positive attributes of the present study, namely: (a) insensitivity to
dust due to the use of near-IR data; (b) clear separation of the
intermediate-age and old stellar populations that are most likely to
trace the underlying mass distribution of the LMC disk; and (c) superb
statistics with approximately a quarter of a million individual
tracers. The present analysis therefore allows us to address the LMC
morphology and its radial dependence in a way that few if any previous
studies could.

Previous authors have generally found the major axis position angle
and ellipticity of tracer distributions in the main body of the LMC to
be in the ranges ${\rm PA}_{\rm maj} = 160^{\circ}$--$190^{\circ}$ and
$\epsilon = 0.1$--$0.3$. These results are not inconsistent with those
presented here, which yield at radii $r \gta 5''$ that ${\rm PA}_{\rm
maj} = 189.3^{\circ} \pm 1.4^{\circ}$ and $\epsilon = 0.199
\pm 0.008$ (cf.~Section~\ref{ss:ellPA}). Figure~\ref{f:ellpar} shows
that the LMC has considerable radial gradients in both its position
angle and ellipticity, so the sizeable differences between the results
from different authors may well be due to the fact that they
studied tracers at different radii. de Vaucouleurs \& Freeman (1973)
found the optical center of the bar to be at RA = $5\chour 23.6\cmin$
and DEC = $-69^{\circ} 44'$ (J2000.0). This agrees with the center
inferred in Section~\ref{ss:center} to within $0.14^{\circ}$, i.e.,
less then three pixels in our map. Bothun \& Thompson (1988)
determined an exponential scale length $r_d = 1.68^{\circ}$ from a
$B$-band surface brightness profile, not dissimilar to the value $r_d
= 1.44^{\circ}$ found in Section~\ref{ss:radprofile}.

Weinberg \& Nikolaev (2000) previously used the same 2MASS data that
are used here to study the structure of the LMC. Their estimates of
the center, ellipticity, position angle and exponential disk scale
length agree in some globally averaged sense with the ones obtained
here. However, a detailed comparison is not possible because they
restricted their analysis to fits of either circular exponential disks
models or spherical power-law models. Such models yield only limited
insight because they have fixed centers, projected ellipticities and
position angles. Instead, it is evident from Figure~\ref{f:ellpar}
that all of these quantities actually have large gradients in the
LMC. In addition, I will show that the LMC is quite non-circular
(Section~\ref{ss:isofaceon} below).

\section{The Deprojected Stellar Number Density}
\label{s:faceon}

In addition to studying the number density distribution in the plane
of the sky, as done in the preceding sections, one can also deproject
the data to obtain the number density distribution in the plane of the
LMC itself.

\subsection{Coordinate Systems}
\label{ss:coordfaceon}

As in Paper~I, I define a cartesian coordinate system $(x',y',z')$ such
that the $(x',y')$-plane contains the LMC disk and the $x'$-axis is
along the line of nodes (the intersection of the $(x',y')$-plane and
the plane of the sky). The LMC plane is inclined with respect to the
sky plane by an angle $i$ (with face-on viewing corresponding to
$i=0$), and the position angle of the line of nodes (measured
counterclockwise from North) is $\Theta$. The near side of the LMC
plane lies at $\Theta_{\rm near} \equiv \Theta - 90^{\circ}$ and the
far side at $\Theta_{\rm far} \equiv \Theta + 90^{\circ}$.

To calculate the position of a star in the $(x',y',z')$ coordinate
system one needs both the position $(\alpha,\delta)$ on the sky and
the distance $D$. However, one does not know the distances to
individual stars. On the other hand, the LMC disk is believed to be
thin (see the discussion in Section~9.3 of Paper~I) so that one can
assume approximately that $z'=0$. Combination of equations~(7) and~(8)
from Paper~I then yields
\begin{eqnarray}
\label{xyprimeforzprimezero}
x' & = & D_0 \> \cos i \sin \rho \cos(\phi-\Theta_{\rm far})
         \> / \>
             [\cos i \cos \rho -
              \sin i \sin \rho \sin(\phi-\Theta_{\rm far} ]  , \nonumber\\
y' & = & D_0 \> \sin \rho \sin(\phi-\Theta_{\rm far})
         \> / \>
             [\cos i \cos \rho -
              \sin i \sin \rho \sin(\phi-\Theta_{\rm far}) ]  ,
\end{eqnarray}
where $D_0$ is the distance to the LMC center. In practice it is
useful not to use the $(x',y')$ coordinate system in the LMC disk
plane, but a new system $(x'',y'')$ that is rotated by an angle
$\Theta_{\rm far}$:
\begin{eqnarray}
\label{rotdouble}
x'' & = & x' \cos \Theta_{\rm far} - y' \sin \Theta_{\rm far} , \nonumber\\
y'' & = & x' \sin \Theta_{\rm far} + y' \cos \Theta_{\rm far} .
\end{eqnarray}
For completeness, I define a coordinate $z''$ that is identical to
$z'$.  With these definitions the line of nodes lies at the same angle
in the $(x'',y'')$ plane of the LMC as in the projected $(X,Y)$ plane
of the sky (defined in Section~\ref{ss:coords}). The coordinates
$(x',y')$ and $(x'',y'')$ all scale linearly with the galaxy distance
$D_0$.

\subsection{Map Creation and Results}
\label{ss:mapfaceon}

For the creation of a face-on map of the LMC I proceeded similarly as
in Section~\ref{s:maps}. The same stars were selected from the DENIS
and 2MASS surveys as before. For each star,
equations~(\ref{xyprimeforzprimezero}) and~(\ref{rotdouble}) were used
to obtain the coordinates $(x'',y'')$ in the plane of the LMC, given
the angular coordinates $(\rho,\phi)$ on the plane of the sky
(calculated from $(\alpha,\delta)$ as in Section~\ref{ss:coords}) and
the viewing angles $(i,\Theta)$ that were determined in Paper~I. The
stars were then binned into square pixels of size $\Delta x'' = \Delta
y'' \equiv \Delta'' = \pi (0.05^{\circ}/180^{\circ}) D_0$. In the
limit of small $\rho$, and along the line-of-nodes, this choice yields
the same pixel size as for the sky maps discussed in
Section~\ref{s:maps}.  For $D_0 = 51 \kpc$ the pixel size is $\Delta''
= 44.5 \pc$. The number of stars in a pixel was divided by its area
$\Delta x'' \> \Delta y''$ to obtain $\mu$, the number density of
stars per unit area. Since it was assumed that all stars have $z'=0$,
whereas in reality the LMC disk has a finite thickness, $\mu$ is best
thought of as the three-dimensional density of the LMC projected
perpendicular to the disk plane, and viewed from that same direction.

As before, maps were smoothed using a Gaussian with a dispersion of 1
pixel to reduce noise. A linear fit to the data in the corners of each
map was used to estimate and subtract the Galactic foreground
contribution. A combined map was created from the 2MASS and the DENIS
data as before, by using the DENIS data (scaled appropriately) to fill
in the regions that are missing from the 2MASS data.

Figure~\ref{f:maps}d shows the resulting face-on map of the LMC.  The
most obvious characteristic of the map is that the LMC is quite
elongated in its disk plane, and is not close to circular at all.
This is further discussed below.

To understand the relation of the face-on map to the sky projection
map shown next to it in Figure~\ref{f:maps}c, note that the line of
nodes (indicated by white line segments) lies at the same angle in
both panels. The sky projection map is obtained conceptually from the
face-on map by rotating the latter around the line of nodes through an
angle $i=34.7^{\circ}$, such that the top left part of the map is
tilted out of the paper.

\subsection{Ellipse Fit Analysis}
\label{ss:isofaceon}

For quantitative interpretation of the face-on deprojected map I used
the same procedure of ellipse fitting as
before. Figure~\ref{f:ellparFO} shows the parameters of the
best-fitting ellipses. The radius $r''$ along the abscissa is the
semi-major axis length of an ellipse in the $(x'',y'')$ coordinates of
the LMC plane defined in Section~\ref{ss:coordfaceon}.

Figure~\ref{f:ellparFO}a shows the radial number density profile. The
dashed line shows the best fitting exponential profile, which has
scale length $r''_d = 1.54 \kpc$. As before, an exponential model does
not provide a very good description of the LMC. In the range $2 \kpc
\lta r'' \lta 4 \kpc$ the surface density shows a pronounced deficit
compared to the best exponential fit, and for $5
\kpc \lta r'' \lta 8 \kpc$ there is an excess number density compared 
to the fit.

Figures~\ref{f:ellparFO}c,e show the radial profiles of the
ellipticity $\epsilon''$ and the major axis position angle ${\rm
PA}''_{\rm maj}$ (measured counterclockwise from the $y''$-axis,
defined modulo $180^{\circ}$). These profiles have the same
qualitative behavior as the quantities $\epsilon$ and ${\rm PA}_{\rm
maj}$ that describe the sky-projection map. At radii $r'' \gta 4 \kpc$
the contour shapes converge to an approximately constant position
angle and ellipticity. The RMS variations in position angle and
ellipticity at $r'' > 5 \kpc$ are $3.7^{\circ}$ and $0.014$,
respectively. The average values and their formal errors\footnote{The
quoted errors do not include the propagated influence of errors in the
adopted LMC viewing angles. Changes in the viewing angles will cause
changes in both ${\rm PA}''_{\rm maj}$ and $\epsilon''$. However, the
errors in the viewing angles are small enough that they do not
influence any of the main results of this paper in a qualitative
sense.} over this radial range are ${\rm PA}''_{\rm maj} =
202.7^{\circ} \pm 1.9^{\circ}$ and $\epsilon'' = 0.312 \pm 0.007$.

The center of a fitted ellipse is denoted
$(x''_0,y''_0)$. Figures~\ref{f:ellparFO}b,d show the radial profiles
of $x''_0$ and $y''_0$. Figure~\ref{f:ellparFO}f shows how
$(x''_0,y''_0)$ drifts in the plane of the LMC as the semi-major axis
length of the contours changes. At radii $r'' \lta 5
\kpc$ the qualitative behavior is similar as for the quantities
$(X_0,Y_0)$ that describe the sky-projection map. However, at larger
radii the behavior is different. In the face-on map the contour
centers in the region $5 \kpc \lta r \lta 8 \kpc$ are all quite close
together (RMS positional difference $0.1 \kpc$), indicating that the
outer contours have more or less a common center in the LMC disk
plane. By contrast, in the sky-projection map the center of the outer
contours drifts to the North-North-East. This can be explained as the
result of viewing perspective, as discussed below.

\section{The Influence of Perspective on the Sky Map}
\label{s:perspective}

Perspective plays an important role in our view of the LMC, due to its
large angular extent on the sky. To illustrate this, consider a set of
ellipses in the $(x'',y'')$ plane of the LMC, centered on the
origin. The ellipses are chosen to be concentric, and to have a fixed
position angle ${\rm PA}''_{\rm maj} = 202.7^{\circ}$ and ellipticity
$\epsilon'' = 0.312$, as appropriate for the outer contours in the LMC
disk plane (cf.~Section~\ref{ss:isofaceon}). Figure~\ref{f:perspec}b
shows the ellipses for a range of semi-major axis lengths. The
discussions in Sections~\ref{ss:coords} and~\ref{ss:coordfaceon}
showed how the $(x'',y'')$ coordinates are obtained from the projected
coordinates $(X,Y)$ on the plane of the sky. The resulting formulae
can be inverted numerically to obtain $(X,Y)$ for given
$(x'',y'')$. Figure~\ref{f:perspec}a shows the sky projection of the
ellipses shown in Figure~\ref{f:perspec}b, obtained using this
inversion.

The projections in the sky plane are well fitted by pure ellipses
without higher-order deviations. The major axis position angle and
ellipticity are ${\rm PA}_{\rm maj} = 193.1^{\circ}$ and $\epsilon =
0.183$, with negligible dependence on the semi-major axis
length. However, the projections are {\it not} concentric on the plane
of the sky, due to the effect of perspective. The near side of an
ellipse in the $(x'',y'')$ plane of the LMC subtends a larger angle on
the sky than the far side. This causes a distortion in the projected
shape that manifests itself as an apparent shift of the center towards
the near side. This is clearly seen in Figure~\ref{f:perspec}c, which
shows the projection of the outermost ellipse in the $(x'',y'')$
plane, together with the major and minor axis of the ellipse that best
describes it. The center is offset from the center of the inner
contours by $0.8^{\circ}$. Figure~\ref{f:perspec}d shows how the
distance $r_{\rm offset}$ by which the center shifts depends on the
semi-major axis length. The offset is in the direction of ${\rm
PA}_{\rm offset} = 26.4^{\circ}$, with almost negligible dependence on
the semi-major axis length. This is between the position angle
$\Theta_{\rm near} = 32.5^{\circ}$ towards the near side of the LMC
(Paper~I) and the position angle ${\rm PA}_{\rm pr-maj} = 20.7^{\circ}$
to which the intrinsic major axis of the LMC projects.  Note that the
latter is not identical to the major axis of the ellipse projections
on the sky, which is at ${\rm PA}_{\rm maj} = 13.1^{\circ}$.
Figure~\ref{f:perspec}e schematically shows the directions of these
variations position angles.

In practice one does not study the projected shape of elliptical rings
in the LMC plane, but the shapes of contours of equal number density.
There is then a second effect that plays a role. A fixed area on the
near side of the LMC disk plane subtends a larger solid angle on the
sky than an area of the same size on the far side of the disk
plane. As a result, the projected number density $\Sigma$ per square
degree is boosted on the far side of the disk relative to the near
side of the disk. This `solid-angle effect' tends to shift the center
of the projected contours of $\Sigma$ in the opposite direction as the
perspective effect discussed above. Which of the two effects dominates
depends on the detailed radial dependence of the number density $\mu$
(per square kpc) in the LMC disk plane. In simple approximation one
may assume that $\mu$ follows an exponential distribution with a scale
length of $r''_d = 1.54 \kpc$ (cf.~Section~\ref{ss:isofaceon}).
Calculations then show that the perspective effect and the solid-angle
effect approximately cancel each other for contours with semi-major
axis length $r \lta 5^{\circ}$. The perspective effect dominates at
large radii, with an expected shift in the center of the projected
number density contours by about $\sim 0.4^{\circ}$ in the direction
of ${\rm PA}_{\rm offset} = 26.4^{\circ}$ when the semi-major axis
length increases from $r = 5^{\circ}$ to $r = 8^{\circ}$.

In practice there is no need to model these effects in approximate
fashion, since the position of each star can be deprojected
individually, as was done in Section~\ref{s:faceon}. However, the
above arguments do provide valuable insight into the details of the
ellipse fits results shown in Figures~\ref{f:ellpar}
and~\ref{f:ellparFO}. In particular, at radii $r \gta 5 \kpc$ the
contour center stays more or less fixed in the deprojected face-on
map. At the same radii, the contour center on the sky drifts in a
direction that is consistent with the value ${\rm PA}_{\rm offset} =
26.4^{\circ}$ predicted by the perspective effect (this direction is
indicated with an arrow in Figure~\ref{f:ellpar}f).

At radii $r \lta 5^{\circ}$ the contour centers drift in a
qualitatively similar way in the projected sky map and the deprojected
face-on map. The size of this drift is considerably more than can be
attributed to the perspective effect, independent of the assumed
viewing angles of the LMC. It must therefore be intrinsic, and hence,
the bar region of the LMC is lopsided, both intrinsically and in
projection. This lopsidedness is due primarily to the faint spiral arm
features named `B1' and `B2' by de Vaucouleurs \& Freeman (1973),
which extend in opposite directions from the western end of the
bar. These features cause an overall movement of the contour centers
in the westward direction for semi-major axis lengths increasing up to
$r \approx 4^{\circ}$, followed by a return in the eastward direction
for larger contours that are not affected by the B1 and B2 features.

It is a generic prediction of models for the LMC that the center of
the outermost number density contours on the sky should drift in the
general direction of the near side of the LMC disk plane. As discussed
in Paper~I, many previous authors have believed that the LMC is
approximately circular with the line of nodes coincident with the
observed major axis position angle ($\Theta = {\rm PA}_{\rm maj}
\approx 190^{\circ}$) and the near side of the LMC plane at position
angle $\Theta_{\rm near} = \Theta - 90^{\circ} \approx 100^{\circ}$.
If this were true, then the center of the outermost number density
contours on the sky should have drifted approximately Eastward, which
is not observed.  While one could assume that the outer LMC is
intrinsically lopsided in such a way as to counterbalance this, Occam's
razor suggests that this is not the preferred solution. Instead, the
observed drift is exactly in the direction suggested by the viewing
angles in Paper~I, and therefore is best interpreted as an independent
confirmation of those viewing angles.

\section{The Elongation of the LMC Disk}
\label{s:elong}

\subsection{The Intrinsic Shapes of Disk Galaxies}
\label{ss:diskshapes}

While the line of nodes for the LMC lies at position angle $\Theta =
122.5^{\circ} \pm 8.3^{\circ}$ (cf.~Paper I), the major axis position
angle of the outer number density contours lies at ${\rm PA}_{\rm maj}
= 189.3^{\circ} \pm 1.4^{\circ}$ (cf.~Section~\ref{ss:ellPA}). The
fact that the major axis doesn't coincide with the line of nodes
implies that the LMC is not intrinsically circular at large radii.
Indeed, deprojection shows that the outer contours have an intrinsic
ellipticity $\epsilon'' = 0.312 \pm 0.007$
(cf.~Section~\ref{ss:isofaceon}).

The conclusion that the LMC is elongated is in itself not surprising.
It is generally expected that disk galaxies are elongated rather than
circular. It is possible to construct self-consistent dynamical models
for elliptical disks (e.g., Teuben 1987), and it is known that bars
and other planar non-axisymmetric structures are common in disk
galaxies. The dark matter halos predicted by cosmological simulations
are generally triaxial (e.g., Dubinski \& Carlberg 1991), and the
gravitational potential in the equatorial plane of such halos does not
have circular symmetry.

The ellipticity of disk galaxies has been a subject of increasing
study in recent years. Studies of the apparent axis ratio distribution
of spiral galaxy disks (Binney \& de Vaucouleurs 1981; Lambas, Maddox
\& Loveday 1992) of individual galaxies (Rix \& Zaritsky 1995;
Schoenmakers, Franx \& de Zeeuw 1997; Kornreich, Haynes \& Lovelace
1998; Andersen \etal 2001) and of the scatter in the Tully-Fisher
relation (Franx \& de Zeeuw 1992) indicate that the average
(deprojected) ellipticity of spiral galaxy is $\epsilon'' \approx
5$--10\%. So while spiral galaxies are generally elongated, their
elongation is usually smaller than inferred here for the LMC. Note
however that Andersen \etal (2001) found two spiral galaxies in their
sample with fairly large elongations ($\epsilon \approx 0.2$);
interestingly, both these galaxies are reported to show faint
interacting companions. Also, it should be kept in mind that galaxies
of type Sm and Im are (by definition) more irregular and lopsided than
spirals (de Vaucouleurs \& Freeman 1973). So it is not a priori clear
whether or not the LMC is atypically elongated for its Hubble type.

\subsection{Evidence for Tidal Effects in the LMC}
\label{ss:tidalevidence}

The most logical explanation for the large elongation of the LMC
appears to be that its shape has been distorted through its tidal
interactions with the Galaxy and/or the SMC, for which there is a
variety of evidence. In its central $4 \kpc$ the LMC has a pronounced
bar, and it is well known that bar formation can be triggered by
interactions (e.g., Gerin, Combes \& Athanassoula 1990; Barnes
1998). The bar is significantly lopsided,
cf.~Figure~\ref{f:ellparFO}f. Observational studies have demonstrated
that lopsidedness is not uncommon in spiral galaxies (Baldwin,
Lynden-Bell \& Sancisi 1980; Rix \& Zaritsky 1995), and that this
lopsidedness is generally caused by lopsidedness of the underlying
mass distribution, and not merely by recent star formation (Rudnick \&
Rix 1998; Swaters \etal 1999). Interactions are believed to be the
dominant mechanism for triggering this lopsidedness (Zaritsky \& Rix
1997). At large radii the Magellanic Stream (e.g., Br\"uns, Kerp \&
Staveley-Smith 2000) and the more recently discovered Leading Arm
(Putman \etal 1998) provide ample evidence for the tidal interaction
between the Milky Way and the Magellanic Clouds (e.g., Gardiner \&
Noguchi 1996).

It is worth noting in this context that the LMC number density profile
shown in Figure~\ref{f:ellparFO}a has an upturn at $r'' \approx 4
\kpc$, beyond which the density shows an excess over the best-fit 
exponential model. This may well be another tidally induced feature of
the LMC. Numerical simulations of satellite tidal disruption naturally
produce upturns that are qualitatively similar to that seen in the LMC
(e.g., Johnston, Spergel, \& Hernquist 1995).

\subsection{The Tidal Influence of the Milky Way}
\label{ss:tidalMW}

Weinberg (2000) has recently stressed the importance of the Galactic
tidal field for the structure of the LMC. He estimates that the tidal
radius of the LMC is $\sim 11 \kpc$, but that at $5 \kpc$ from the
center the ratio of the tidal force to the self-force has dropped by
only $\sim 20$\%. Tidal effects are therefore expected to influence
LMC structure well inside the tidal radius. Weinberg stressed
primarily the fact that vertical thickening of the disk results from
tidal forces, but one would also expect an in-plane elongation to be
induced. Quantitative estimates for the induced elongation are not
readily available. However, to lowest order one would expect the main
body of the LMC to become elongated in the direction of the tidal
force, i.e., towards the Galactic Center. Material that is tidally
stripped will phase mix along the orbit (e.g., Johnston, Hernquist \&
Bolte 1996). For an orbit that is not too far from being circular, the
direction along the orbit is perpendicular to the direction towards
the Galactic Center. Hence, one expects the elongation of the main
body to be perpendicular to any tidal streams that emanate from it.
This generic feature of satellite disruption is often seen both in
numerical simulations (e.g., figure~3 in Johnston 1998; figure~2 in
Helmi \& White 2001) and in observations (e.g., figure~3 of
Odenkirchen \etal 2001, for the case of the globular cluster Pal~5).
Of course, in reality matters are complicated for the LMC by the fact
that its orbit is not circular (Weinberg (2000) estimates that the
ratio of the apogalactic and perigalactic distance is $\sim 2:1$, with
the LMC currently being near pericenter) and by the fact that the
Galactic Center does not lie in the plane of the LMC disk (see
below). Nonetheless, it appears reasonable to assume that one should
generically expect in any model in which the LMC elongation is induced
by the tidal force of the Milky Way that: (a) the disk elongation
should point roughly towards the Galactic Center (or its projection
onto the disk plane); and (b) the disk elongation should be roughly
perpendicular to any tidal streams from the LMC (or their projection
onto the disk plane).

Figure~\ref{f:schemview} shows a schematic representation of the main
features of the Magellanic system as seen on the projected $(X,Y)$
plane of the sky. The LMC is represented as an ellipse with the size,
ellipticity and position angle of the outermost contour in the near-IR
map shown in Figure~\ref{f:maps}c. The proper motion vectors for the
LMC and the SMC show that these galaxies move in more or less parallel
directions (Kroupa \& Bastian 1997). The Magellanic Stream trails the
LMC and the SMC, and as expected, is aligned roughly parallel with
their proper motions. The Leading Arm makes an angle with the
Magellanic Stream in agreement with model predictions for the tidal
origin of both features (Gardiner \& Noguchi 1996). On the sky, as
seen from the center of the LMC, the Galactic Center lies at a
position angle of $183.7^{\circ}$ (at a distance $\rho = 81.5^{\circ}$
from the LMC). The outer contours of the LMC near-IR starcount map
have major axis position angle ${\rm PA}_{\rm maj} = 189.3^{\circ} \pm
1.4^{\circ}$, cf.~Section~\ref{ss:ellPA}. The projected elongation of
the LMC is therefore aligned to within $\sim 6^{\circ}$ with the
projected direction to the Galactic Center. The proper motion of the
LMC is in the direction of position angle $100^{\circ} \pm 5^{\circ}$,
which is also more or less the direction of the Magellanic
Stream. This is perpendicular to ${\rm PA}_{\rm maj}$ within the
errors. Given the above arguments, these results are consistent with
the hypothesis that the elongation of the LMC is induced by the tidal
force of the Milky Way.

More comprehensive insight can be obtained from a three-dimensional
analysis. To this end I use the coordinate system $(x'',y'',z'')$
defined in Section~\ref{ss:coordfaceon}, where the $(x'',y'')$ plane
is the plane of the LMC disk. The positions of the LMC, the SMC and
the Galactic Center can be calculated in this coordinate system with
the help of equation~(\ref{rotdouble}) and the formulae in Paper~I,
using the known distances $D_{\rm LMC}
\approx 51 \kpc$, $D_{\rm SMC} \approx 62 \kpc$ (e.g., Cioni \etal 2000c) 
and $D_{\rm GC} \approx 8 \kpc$. Figure~\ref{f:schemviewthreeD}a shows
the $(x'',y'')$ projection of the three-dimensional $(x'',y'',z'')$
space. The Galactic Center lies at position angle ${\rm PA}'' = 42.0$,
measured counterclockwise from the $y''$-axis. The outer contours of
the LMC near-IR starcount map have major axis position angle ${\rm
PA}''_{\rm maj} = 22.7^{\circ} \pm 1.9^{\circ}$ (defined modulo
$180^{\circ}$), cf.~Section~\ref{ss:isofaceon}. Thus, the elongation
of the LMC is aligned to within $\sim 21^{\circ}$ with the direction
to the projection of the Galactic Center onto the disk plane. While
this alignment is less accurate than the projected alignment on the
sky (which is therefore somewhat misleading), the alignment is still
better than what one would expect in a randomly drawn situation. The
LMC velocity vector in the $(x'',y'')$ plane lies at position angle
${\rm PA}'' = 116^{\circ} \pm 5^{\circ}$. As in the sky-projection
view of Figure~\ref{f:schemview}, this is perpendicular to ${\rm
PA}''_{\rm maj}$ within the errors. If one assumes that the direction
of the Magellanic Stream follows the velocity vector (as it does in
projection on the sky), then the disk elongation is perpendicular to
the direction of the Magellanic Stream as projected onto the LMC disk
plane. As before, these results are consistent with the hypothesis
that the elongation of the LMC is induced by the tidal force of the
Milky Way.

Figure~\ref{f:schemviewthreeD}b shows the $(x'',z'')$ projection of
the three-dimensional $(x'',y'',z'')$ space. The Galactic Center does
not lie in the plane of the LMC, but lies at an angle of
$62.9^{\circ}$ from the plane as seen from the LMC center. The
component of the tidal force parallel to the LMC disk plane is
therefore considerably less than the total tidal force. However, the
calculations of Weinberg (2000) show that the LMC disk is expected to
precess and nutate through large angles as it moves around the
Galaxy. The angular distance of the Galactic Center from the LMC disk
plane therefore varies as function of position along the orbit, and
the average tidal force parallel to the LMC disk plane may be larger
than what it currently is.

\subsection{The Tidal Influence of the SMC}
\label{ss:tidalSMC}

The tidal force due to a massive body is proportional to its mass
divided by the cube of the distance. The mass of the SMC is $\sim 3
\times 10^9 \Msun$ (Gardiner \& Noguchi 1996) and the mass of the
Milky Way enclosed within the LMC orbit is $\sim 5 \times 10^{11}
\Msun$ (Kochanek 1996). The positions in the $(x'',y'',z'')$ coordinate 
system shown in Figure~\ref{f:schemviewthreeD} yield $D_{\rm LS} =
23.7 \kpc$ for the distance between the LMC and the SMC and $D_{\rm
LG} = 50.4\kpc$ for the distance between the LMC and the Galactic
Center. As a result, the tidal force from the Milky on the LMC is $17$
times larger than that from the SMC. So this would appear to indicate
that the tidal effect of the SMC on the LMC is negligible compared to
the that of the Milky Way. However, it cannot be ruled out that the
LMC and the SMC had a significant interaction in the past.  This
depends sensitively on the poorly known relative orbit of these
galaxies. It may even be that the SMC originated in a tidal disruption
event of the LMC. If the SMC were ever as close as $\sim 10
\kpc$ from the center of the LMC, essentially touching the outer
contours of the LMC, its tidal influence would definitely have
exceeded that of the Milky Way.

On the projected plane of the sky (Figure~\ref{f:schemview}), as seen
from the center of the LMC, the SMC lies at a position angle of
$228.9^{\circ}$ (at a distance $\rho = 21.2^{\circ}$ from the
LMC). The outer contours of the LMC near-IR starcount map have major
axis position angle ${\rm PA}_{\rm maj} = 189.3^{\circ} \pm
1.4^{\circ}$, cf.~Section~\ref{ss:ellPA}. The projected elongation of
the LMC is therefore aligned only to within $\sim 40^{\circ}$ with the
projected direction to the SMC. On the other hand, the alignment is
better in the $(x'',y'')$ plane of the LMC (see
Figure~\ref{f:schemviewthreeD}a). The SMC lies at position angle ${\rm
PA}'' = 228.7$, measured counterclockwise from the $y''$-axis. The
outer contours of the LMC near-IR starcount map have major axis
position angle ${\rm PA}''_{\rm maj} = 202.7^{\circ} \pm 1.9^{\circ}$,
cf.~Section~\ref{ss:isofaceon}.  Thus, the elongation of the LMC is
aligned to within $\sim 26^{\circ}$ with the direction to the
projection of the SMC onto the disk plane.  This is only marginally
worse than the alignment with respect to the projected position of the
Galactic center. In addition, the SMC lies close to the plane of the
LMC disk (see Figure~\ref{f:schemviewthreeD}b), at an angle of
$15.6^{\circ}$ as seen from the LMC center. So currently, most of the
tidal force of the SMC acts parallel to the LMC disk plane.

Upon assessing all the above arguments, it appears most likely that
the elongation of the LMC is due to the tidal force from the Milky
Way, and not that from the SMC. This is suggested in particular by the
facts that the elongation of the LMC disk is perpendicular to the
Magellanic Stream, and that the tidal force from the Milky Way
currently exceeds that from the SMC by a large factor.

\section{Kinematics of the LMC}
\label{s:kinematics}

Many previous investigations have studied the kinematics of tracers in
the LMC disk. The results have generally been interpreted under the
assumption that the LMC disk is circular, and that the tracers move on
circular orbits. In such models the line of maximum gradient in the
velocity field coincides with the line of nodes, and this method has
therefore often been used to estimate the viewing geometry of LMC, as
reviewed in Paper~I. However, as demonstrated here, the LMC disk is
not circular but elongated. It is therefore worthwhile to rediscuss
the observed kinematics in light of this finding.
 
\subsection{Gas kinematics}
\label{ss:gaskin}

The gas kinematics of the LMC have been studied primarily using HI
(e.g., Rohlfs \etal 1984; Luks \& Rohlfs 1992; Kim \etal 1998). Kim
\etal (1998) analyzed data from the Australia Telescope Compact Array
HI survey using a standard circular tilted-ring algorithm. At the
outermost available radii in their study ($\rho =
2.9^{\circ}$--$3.5^{\circ}$) they obtained $\Theta_{\rm max} = 158 \pm
2^{\circ}$ for the position angle of the line of maximum velocity
gradient. This differs considerably from the position angle of the
line of nodes inferred in Paper~I, $\Theta = 122.5 \pm 8.3^{\circ}$.

In an elliptical disk one does not expect $\Theta_{\rm max}$ and
$\Theta$ to be equal. The kinematical properties of an elliptical gas
disk can be approximated using epicyclic theory (e.g., Gerhard \&
Vietri 1986; Franx, van Gorkom \& de Zeeuw 1994; Schoenmakers \etal
1997). The cold gas is expected to move along stable closed orbits in
a symmetry plane (Tohline, Simonson \& Caldwell 1982). In general the
predicted velocity field depends on many unknown factors, such as the
ellipticity of the gravitational potential in the disk plane, the
radial slope of the circular velocity curve, and whether or not
gravitational potential is tumbling or not. However, analytical
results can be obtained in the simple case in which the gravitational
potential is stationary, and corresponds to a flat rotation curve
(Franx \etal 1994). One then obtains
\begin{equation}
  \Theta_{\rm max} - \Theta = (180/\pi)
      \epsilon_{\rm gas} \sin 2\phi_{\rm gas} \cos i , \qquad
  \Theta_{\rm 0} - \Theta   = 90 + (180/\pi)
      \epsilon_{\rm gas} \sin 2\phi_{\rm gas} / \cos i , \qquad
\label{gasangles}
\end{equation}
where $\Theta_{\rm 0}$ is the position angle of the line of zero
line-of-sight velocity, the angles on the left hand side are in
degrees, $\epsilon_{\rm gas}$ is the ellipticity of the gas disk, and
$\phi_{\rm gas}$ is the angle in the plane of the gas disk from its
major axis to the line of nodes. The equations are correct only to
lowest order in $\epsilon_{\rm gas}$, and therefore formally apply
only to nearly-circular disks.

Use of equations~(\ref{gasangles}) is not straightforward in the
context of the LMC, since the HI gas distribution is very filamentary
(Kim \etal 1998), and it is not easy to assign a unique ellipticity
and major axis position angle. So let us consider merely what the
maximum misalignment is that one could possibly expect, obtained when
$\sin 2\phi_{\rm gas} = \pm 1$. This yields, with the inclination
angle from Paper~I, that $\Theta_{\rm max} - \Theta = \pm 47.1
\epsilon_{\rm gas}{}^{\circ}$ and $\Theta_0 - \Theta = 90^{\circ} \pm 69.7
\epsilon_{\rm gas}{}^{\circ}$. The line of maximum velocity 
gradient and the line of zero velocity are not perpendicular; in this
maximum misalignment case $\Theta_0 - \Theta_{\rm max} = 90^{\circ}
\pm 22.6 \epsilon_{\rm gas}{}^{\circ}$. The observed difference
$\Theta_{\rm max} - \Theta = 35.5^{\circ} \pm 8.5^{\circ}$ (using the
Kim \etal 1998 data at radii $\rho = 2.9^{\circ}$--$3.5^{\circ}$)
would clearly not generally be expected in the simple model described
above unless the gas disk had an unrealistic elongation,
$\epsilon_{\rm gas} = 0.75$, and even then only in the most favorable
situation ($\sin 2\phi_{\rm gas} = 1$).

Although the observed misalignment between the line of maximum HI
velocity gradient and the line of nodes cannot be easily explained
with a simple model of non-circular motion in an elliptical disk, the
HI data presented by Kim \etal (1998) certainly do not support the
idea that the gas disk would be in circular motion. The value of
$\Theta_{\rm max}$ shows a statistically significant twist of $\sim
20^{\circ}$ from small to large radii, indicating that the HI disk
cannot have an intrinsically constant ellipticity and position angle
(not surprising, given that this is not true for the stellar disk
either; cf.~Figure~\ref{f:ellparFO}). The lines of maximum velocity
gradient and of zero velocity are not perpendicular, as generically
expected in a non-circular disk. The inclination derived from the
shape of the HI brightness contours under the assumption of circular
symmetry is inconsistent with the value derived from fits to the
velocity field, which also argues for a non-circular gas disk. In
addition, the gas disk may not even be close to equilibrium. The
center of the HI rotation velocity field (RA = $5\chour 17.6\cmin$ and
DEC = $-69^{\circ} 02'$, Kim \etal 1998) does not coincide with either
the center of the bar, or the center of the outer contours of the
stellar distribution. It is offset from both by $\sim 1 \kpc$, as
indicated in Figures~\ref{f:ellpar}f and~\ref{f:ellparFO}f. Also, the
LMC and the SMC are enshrouded in a common HI envelope, and they are
connected by a bridge of HI gas. This bridge connects to the LMC disk
at a distance of only $\sim 3.3^{\circ}$ from the LMC center (Putman
\etal 1998). Consequently, even at small radii the LMC gas disk
appears to be subject to tidal disturbances that probably also affect
the velocity field. This makes it unlikely that any simple equilibrium
model for the gas kinematics, such as that leading to
equation~(\ref{gasangles}), may provide much insight into the
structure of the LMC.

\subsection{Kinematics of Discrete tracers}
\label{ss:starkin}

Not only HI gas, but also a variety of discrete tracers have been used
for kinematical studies of the LMC. These include star clusters
(Freeman \etal 1983; Schommer \etal 1992), planetary nebulae
(Meatheringham \etal 1988), HII regions and supergiants (Feitzinger
\etal 1977), and carbon-rich AGB stars (Kunkel \etal 1997; Graff \etal
2000; Alves \& Nelson 2000). The line of maximum velocity gradient has
generally been found to be in the range $\Theta_{\rm max} =
140^{\circ}$--$190^{\circ}$. The large variation between different
studies can be attributed in part to the small number statistics of
some studies. The most recent and accurate determinations have come
from studies of carbon stars. Alves \& Nelson (2000) find
that\footnote{The position angles $\Theta_0$ in Table~1 of Alves \&
Nelson (2000) are measured from North over East, in the usual
astronomical convention (Alves, priv.~comm.).} $\Theta_{\rm max} =
143^{\circ} \pm 7^{\circ}$ for carbon stars at radii $2.8^{\circ} \leq
\rho \leq 5.6^{\circ}$. This is roughly the same radial range used in
Paper~I to determine the position angle of the line of nodes, which
yielded $\Theta = 122.5 \pm 8.3^{\circ}$. However, Alves \& Nelson
(2000) find that $\Theta_{\rm max}$ increases with distance from the
center of the LMC, and reaches $\Theta_{\rm max} = 183^{\circ} \pm
8^{\circ}$ in their outermost ring, for which the mean distance to the
LMC center is $9.2^{\circ}$. Hence, $\Theta_{\rm max} -
\Theta$ increases from $20^{\circ} \pm 11^{\circ}$ at intermediate radii to
$60^{\circ} \pm 12^{\circ}$ in the outer parts of the LMC disk. Graff
\etal (2000) find from a study of carbon stars that $\Theta_{\rm max}
= 160^{\circ}$ averaged over all radii, consistent with the results of
Alves \& Nelson (2000).

For a gas disk one can assume that the gas moves on closed orbits,
since shocks rapidly depopulate self-intersecting orbits (Tohline
\etal 1982). This leads to simple formulae such as those in
equation~(\ref{gasangles}). By contrast, discrete tracers behave as a
collisionless population, and can populate any of a large number of
rosetta orbits. Hence, there is no simple method to predict the
misalignment between the line of nodes and the line of maximum
velocity gradient for discrete tracers in an elliptical disk. However,
if one uses equation~(\ref{gasangles}) as a rough guide, it doesn't
seem a priori impossible get a misalignment of $20^{\circ}$ in a disk
with an elongation of $\epsilon'' = 0.31$. On the other hand, it seems
hard to believe that a misalignment as large as $60^{\circ}$ could be
generated naturally as a result of rotation in a somewhat elongated
disk. In view of this it seems worthwhile to investigate whether there
could be any systematic uncertainties or errors in previous
interpretations and modeling of the observed LMC velocity field.

\subsection{Correction for the Center of Mass Motion of the LMC}
\label{ss:cenmass}

One complication in the interpretation of the velocity field of the
LMC arises from the motion of its center of mass. As one moves away
from the LMC center, the transverse component of this motion ceases
being perpendicular to the line of sight. This causes a spurious
solid-body rotation component in the observed line-of-sight velocities
of the form $\Delta v_{\rm los} = v_t \sin \rho \cos (\Phi -
\Theta_t)$, where $\rho$ is the angular distance from the LMC center,
$\Phi$ is the position angle on the sky, and $v_t$ and $\Theta_t$ are
the size and position angle of the transverse velocity vector.  This
spurious solid-body component yields an angular offset between the
true and observed position angle of the line of maximum velocity
gradient. Initially this effect was used to estimate $v_t$ under the
assumption (appropriate for a circular disk) that the line of maximum
velocity gradient should be equal to the photometric major axis
(Feitzinger \etal 1977; Meatheringham \etal 1988). More recently,
actual measurements of the transverse motion of the LMC have become
available from proper motion studies. The average from a variety of
studies as quoted in Kroupa \& Bastian (1997) is $v_t = 389 \pm 46
\kms$ and $\Theta_t = 92^{\circ} \pm 2^{\circ}$ (note that these
values are not corrected for the reflex motion of the sun, by contrast
to those shown in Figures~\ref{f:schemview}
and~\ref{f:schemviewthreeD}).

Most recent studies of the velocity field of the LMC disk, including
e.g.~the HI study by Kim \etal (1998) and the carbon star study by
Alves \& Nelson (2000) discussed above, corrected their data for the
spurious component $\Delta v_{\rm los}$ using existing proper motion
results. However, it should be kept in mind that this correction is
somewhat uncertain. The observed rotation curve, corrected for the
spurious component, peaks at $4 \kpc$ from the LMC center, and then is
approximately flat at $\sim 37 \kms$ (Alves \& Nelson 2000). By
contrast, at $\rho = 9.2^{\circ}$, the mean distance to the LMC center
for the outermost ring in the study of Alves \& Nelson (2000), $\Delta
v_{\rm los}$ peaks at $62 \kms$. So in the outer parts of the LMC the
spurious component in the velocity field exceeds the intrinsic
velocity field by almost a factor $1.7$. The inferred properties of
the intrinsic velocity field at large radii are therefore very
sensitive to small errors in the transverse motion quantities $v_t$
and $\Theta_t$. This is important, since some measurements of these
quantities are quite inconsistent with the results quoted in Kroupa \&
Bastian (1997). For example, Anguita, Loyola \& Pedreros (2000) find
$v_t = 812 \pm 73 \kms$ and $\Theta_t = 30^{\circ} \pm
4^{\circ}$. Also, the motion measured by Kroupa \& Bastian (1997) is
smaller than predicted by models for the origin of the Magellanic
Stream (Lin, Jones \& Klemola 1995). If $v_t$ is larger in reality
than inferred by Kroupa \& Bastian (1997), then the inferred values of
$\Theta_{\rm max}$ would have to be revised downward (Feitzinger \etal
1977), which would bring the observed values more in line with the
position angle $\Theta$ of the line of nodes inferred in Paper~I. So
it is not inconceivable that some or all of the discrepancies between
$\Theta_{\rm max}$ and $\Theta$ discussed in Sections~\ref{ss:gaskin}
and~\ref{ss:starkin} are due to uncertainties in the correction for
the center-of-mass motion of the LMC.

\subsection{Precession and Nutation of the LMC Disk}
\label{ss:precesnut}

Previous authors have interpreted the velocity field of the LMC as a
sum of two components, the motion of the center of mass of the LMC
around the galaxy, and a differential rotation of the LMC disk in its
plane. However, there are almost certainly other components of motion
in the center of mass system. In particular, the N-body calculations
of Weinberg (2000) show that the disk of the LMC is expected to
precess and nutate as the LMC moves along its orbit. This is due to
torquing by the time-variable tidal force from the Milky Way. It is of
interest to assess whether the expected wobbling of the disk might
form a significant component of the observed velocity field. Figures~3
and~4 of Weinberg (2000) show for his calculations the variations in
the azimuthal and polar angles of the vector normal to the disk plane.
The rate of change of the three-dimensional angle is largest near the
end of his simulation (time $= 5.8 \times 10^9$ Gyr), when it is $\sim
340^{\circ}$ per Gyr. If this angular velocity is seen along the line
of sight at $\rho = 9.2^{\circ}$, the mean distance to the LMC center
for the outermost ring in the study of Alves \& Nelson (2000), it
corresponds to a line of sight velocity of $47.6
\kms$. This exceeds the observed velocities, which peak at $\sim 37
\kms$ (Alves \& Nelson 2000). This is the most extreme disk-wobbling 
velocity that can be obtained with Weinberg's calculations, and on
average the precession/nutation velocity may well be less by a factor
of a few. In addition, this velocity need not be aligned with the line
of sight. Nonetheless, it is clear that precession and nutation could
have a significant influence on the observed velocity field. So it is
not inconceivable that some or all of the discrepancies between
$\Theta_{\rm max}$ and $\Theta$ discussed in Sections~\ref{ss:gaskin}
and~\ref{ss:starkin} are due to possible precession and nutation of
the LMC disk plane, which has not been accounted for in any previous
analysis.

\section{Conclusions}
\label{s:conc}

I have used the data from the near-IR 2MASS and DENIS surveys to
create star count maps of the LMC on the projected plane of the sky.
These data are perfectly suited for a study of the structure of the
LMC, because of the large statistics and insensitivity of near-IR data
to dust absorption. The analysis is restricted to RGB and AGB stars
selected from their position in the near-IR CMDs. The limiting
magnitude that is applied to the sample is corrected for the
inclination of the LMC, to avoid distance-induced biases. Galactic
foreground stars make a small contribution to the maps that can be
adequately subtracted. The 2MASS map has some small regions with no
data in the 2MASS Second Incremental Data Release, and the DENIS map
shows some minor residual artifacts of the DENIS scan
pattern. However, the maps from the separate surveys can be combined
into a single map that is free from any obvious blemishes or
artifacts.

The RGB and AGB stars selected for the star count map are part of the
intermediate-age and old stellar populations that are expected to
trace the underlying mass distribution of the LMC disk. The map
clearly shows the well-known bar of the LMC, but is otherwise
remarkably smooth. This contrasts sharply with optical images of the
LMC which show pronounced star forming regions, HII regions, and
spatial variations in dust absorption. There are small hints of spiral
structure, but only at a very low level. There is certainly no well
organized spiral pattern in the LMC. The LMC disk can be traced out to
$r \approx 9^{\circ}$ from the center.

Ellipse fitting can be used for quantitative analysis of the star
count map. The radial number density profile to lowest-order follows
an exponential profile with a scale-length $r_d \approx 1.3
\kpc$. However, there are clear deviations from an exponential profile
at large radii, where there is an excess of stars over the
best-fitting exponential. This may be due to the tidal effect of the
Milky Way on the LMC. The position angle and ellipticity profile both
show large variations as function of radius, but converge to
approximately constant values for $r \gta 5^{\circ}$. In the outer
parts, ${\rm PA}_{\rm maj} = 189.3^{\circ} \pm 1.4^{\circ}$ and
$\epsilon = 0.199 \pm 0.008$. These results are consistent with
previous studies of the distributions of tracers in the LMC disk, but
most of those were of lower accuracy than the results presented here.

The center of the star count contours on the sky shows considerable
variations as function of radius. At small radii $r \lta 5^{\circ}$,
this is due to intrinsic lopsidedness of the LMC: the contours are not
centered on the center of the bar. At large radii the effect of
viewing perspective (one side of the LMC being closer to us than the
other) is the dominant source of drift in the contour center. One
expects on the basis of simple geometry that the contour center should
drift in the direction of the near side of the LMC plane. Indeed, at
large radii the data show a drift in the contour center that is fully
consistent with the LMC viewing angles inferred in Paper~I. This
provides an important, independent confirmation of the latter.

The line of nodes position angle inferred in Paper~I, $\Theta =
122.5^{\circ} \pm 8.3^{\circ}$, differs by as much as $67^{\circ} \pm
8^{\circ}$ from the major axis position angle of the outer LMC star
count contours. This implies that the LMC is not intrinsically
circular at large radii. To study the intrinsic structure of the LMC
one can deproject the data using the viewing angles from Paper~I. The
best-fitting exponential disk scale length is now $r_d \approx 1.5
\kpc$, but again, there are considerable deviations from an
exponential profile at large radii. The position angle and ellipticity
again converge to approximately constant values at large radii. The
intrinsic ellipticity of the LMC disk at large radii is $\epsilon'' =
0.312 \pm 0.007$. The outer contours have a more-or-less common
center, which lies $\sim 0.4 \kpc$ from the center of the bar. Neither
agrees with the center of rotation of the HI gas disk, which is offset
by $\sim 0.8 \kpc$ from both.

The ellipticity of the LMC disk is considerably larger than the
typical ellipticities of disk galaxies, which are believed to be in
the range $0.05$--$0.10$. This suggests that the elongation of the LMC
disk is the result of tidal forces. The Milky Way is the most likely
cause, given that its tidal force is currently $\sim 17$ times larger
than that of the SMC. Indeed, Weinberg (2000) has recently stressed
the large effect that the tidal force of the Milky Way is expected to
have on LMC structure. An analysis of the projected structure of the
Galaxy-LMC-SMC triple system on the sky shows that the projected
elongation of the LMC is aligned to within $\sim 6^{\circ}$ with the
projected direction to the Galactic Center. The elongation is
perpendicular to the LMC proper motion and the Magellanic Stream to
within the errors. A three-dimensional analysis shows that the
Galactic Center lies at an angle of $62.9^{\circ}$ from the plane of
the LMC, as seen from the LMC center. The projection of the Galactic
Center onto the disk plane is aligned to within $\sim 21^{\circ}$ with
the major axis of the disk. The elongation is perpendicular to within
the errors to the projection of the three-dimensional LMC center of
mass velocity vector onto the LMC disk plane. These results are
qualitatively consistent with the hypothesis that the elongation of
the LMC disk is due to the tidal force from the Milky
Way. Quantitative predictions for the induced LMC disk elongation
currently do not exist.

The kinematics of the LMC have been studied previously using both HI
gas and discrete tracers. In the past, the position angle $\Theta_{\rm
max}$ of the line of maximum gradient in the velocity field has
generally been used to estimate the position angle $\Theta$ of the
line of nodes, which assumes that the LMC disk is circular. However,
the results of the present investigation show that the LMC disk is not
circular, and as result, $\Theta$ and $\Theta_{\rm max}$ need not
agree. It is found that $\Theta_{\rm max} - \Theta$ is somewhere in
the range $20^{\circ}$--$60^{\circ}$, depending on the type or tracer
and the radial distance to the LMC center. While values at the lower
end of this range can conceivably be induced by non-circular motions
in a slightly elongated disk, it is hard to believe that this could
cause misalignments as large as $60^{\circ}$. However, the
misalignment may be due to velocity components other than rotation of
tracers in the disk plane itself, namely: (a) uncertainties in the
space of motion of the LMC center of mass; and (b) possible motion of
the LMC plane in an inertial frame attached to its center of mass. The
former causes a spurious solid body component in the observed velocity
field that is generally corrected for. However, the correction is
large, and the space motion of the LMC is not known very
accurately. So the corrections that are generally made may be in
error. The latter has not been taken into account in any study of the
LMC kinematics, but is predicted to occur naturally as the disk plane
precesses and nutates when it moves around the galaxy (Weinberg
2000). In either case there could be a solid body component in
observed line-of-sight velocity fields that has not been corrected
for, and is large enough to cause the difference between $\Theta$ and
the inferred values of $\Theta_{\rm max}$. Note that the LMC has an
approximately flat rotation curve, whereas solid body rotation
increases linearly with radius. So if this is the correct explanation,
then the difference between $\Theta$ and $\Theta_{\rm max}$ should
increase as one moves further out into the LMC, exactly as observed
(Alves \& Nelson 2000).

The main result of the present paper is that the structure of the LMC
disk is considerably more complicated than has been assumed
previously. The LMC disk is not circular at larger radii, tidal
effects may have considerably distorted its shape, and previously
unmodeled velocity components may be contributing to the observed
velocity field. These results should serve as a warning to any
attempts at modeling the LMC as a simple equilibrium system. This may
be particularly important in the context of microlensing studies,
where the structure of the LMC features prominently in ongoing
discussions over the contribution of LMC self-lensing to the inferred
microlensing event rates (e.g., Sahu 1994; Weinberg 2000; Alcock \etal
2000).


\acknowledgments

The analysis made use of data products from the Two Micron All Sky
Survey, which is a joint project of the University of Massachusetts
and the Infrared Processing and Analysis Center/California Institute
of Technology, funded by the National Aeronautics and Space
Administration and the National Science Foundation. I thank Maria Rosa
Cioni for providing the DENIS Catalog towards the Magellanic Clouds
and for comments on a draft of the paper, and David Alves for useful
discussions. The anonymous referee provided useful comments that
helped improve the presentation of the paper.

\clearpage




\ifsubmode\else
\baselineskip=10pt
\fi


\clearpage


\ifsubmode\else
\baselineskip=14pt
\fi


\newcommand{\figcapcmds}
{CMDs for the LMC region of the sky. {\bf (a)} The $(J-K_s,K_s)$ CMD
from 2MASS data. {\bf (b)} The $(I-J,J)$ CMD from DENIS data. In both
panels only a subset of the catalog data is shown, to avoid saturation
of the grey scale. The solid lines in each panel mark the boundary of
the region in which stars were extracted to create LMC number density
maps. The stars in these regions are in large majority RGB and AGB
stars in the LMC.\label{f:cmds}}

\newcommand{\figcapmaps}
{The first three panels show the surface number density distribution
on the sky of RGB and AGB stars in the LMC. North is to the top and
east is to the left. Each panel is $23.55^{\circ} \times
21.55^{\circ}$. The Galactic foreground contribution was
subtracted. {\bf (a; top left)} Stars from the 2MASS survey that fall
in the $(J-K_s,K_s)$ CMD region shown in Figure~\ref{f:cmds}a. The two
dark vertical rectangles are regions missing from the 2MASS Second
Incremental Data Release. {\bf (b; top right)} Stars from the DENIS
survey that fall in the $(I-J,J)$ CMD region shown in
Figure~\ref{f:cmds}b. The features along lines of constant declination
(which run close to vertical) are artifacts along a number of
individual DENIS scan strips. {\bf (c; bottom left)} The same 2MASS
surface density map as in panel (a), but with the unobserved regions
filled in using the DENIS data in panel (b). {\bf (d; bottom right)}
The face-on view of the LMC, deprojected using the viewing angles
determined in Paper~I and the approach described in
Section~\ref{s:faceon}. The image is $20.95 \times 19.18 \kpc$. The
LMC disk is not circular in the disk plane, but is instead
considerably elongated. The line of nodes is indicated by white line
segments; it lies at the same angle in the other panels of the
figure.\label{f:maps}}
 
\newcommand{\figcapellpar}
{Parameters of ellipse fits to the LMC number density distribution on
the sky shown in Figure~\ref{f:maps}c, as function of semi-major axis
length $r$. {\bf (a; top left)} Number density $\Sigma$, in units of
stars per square degree. The number of stars refers to the region of
the $(J-K_s,K_s)$ CMD shown in Figure~\ref{f:cmds}a. The dashed line
is the best-fitting exponential number density profile, which has $r_d
= 1.42^{\circ}$. The dotted horizontal line indicates the average
Galactic foreground contribution in this part of the CMD (which was
subtracted from the map). {\bf (b; top right)} The projected
coordinate $X_0$ of the ellipse center. {\bf (c; middle left)} The
position angle ${\rm PA}_{\rm maj}$.  {\bf (d; middle right)} The
projected coordinate $Y_0$ of the ellipse center. {\bf (e; bottom
left)} The ellipticity $\epsilon$. {\bf (f; bottom right)} The drift
of the ellipse center $(X_0,Y_0)$ in the projected sky plane. Data
points are shown as crosses that correspond to the size of the formal
errors. Consecutive radii are connected by dotted lines. The solid dot
is the position of maximum number density (i.e., the center of the
bar). The arrow indicates the direction of position angle
$26.4^{\circ}$. This is the direction in which the center of the outer
contours is expected to drift due to the effect of viewing
perspective, as discussed in Section~\ref{s:perspective}, given that
the near side of the LMC is at position angle $\Theta_{\rm near} =
32.5^{\circ}$ (cf.~Paper~I). The solid square at the right edge of the
panel indicates the position of the HI rotation center of the LMC,
from Kim \etal (1998).\label{f:ellpar}}

\newcommand{\figcapellparFO}
{Parameters of ellipse fits to the face-on deprojected LMC number
density map shown in Figure~\ref{f:maps}d, as function of semi-major
axis length $r''$. {\bf (a; top left)} Number density $\mu$, in units
of stars per square kiloparsec. The number of stars refers to the
region of the $(J-K_s,K_s)$ CMD shown in Figure~\ref{f:cmds}a. The
dashed line is the best-fitting exponential number density profile,
which has $r''_d = 1.54 \kpc$. The dotted horizontal line indicates
the average Galactic foreground contribution in this part of the CMD
(which was subtracted from the map). {\bf (b; top right)} The
coordinate $x''_0$ of the ellipse center. {\bf (c; middle left)} The
position angle ${\rm PA}''_{\rm maj}$.  {\bf (d; middle right)} The
projected coordinate $y''_0$ of the ellipse center. {\bf (e; bottom
left)} The ellipticity $\epsilon''$. {\bf (f; bottom right)} The drift
of the ellipse center $(x''_0,y''_0)$ in plane of the LMC disk. Data
points are shown as crosses that correspond to the size of the formal
errors. Consecutive radii are connected by dotted lines. The solid dot
is the position of maximum number density (i.e., the center of the
bar). The solid square near the right edge of the panel indicates the
position of the HI rotation center of the LMC, from Kim \etal (1998).
\label{f:ellparFO}}

\newcommand{\figcapperspec}
{Illustration of the effect of perspective on the LMC morphology.
{\bf (a; top left)} projected shapes on the $(X,Y)$ plane of the sky
for the set of concentric ellipses shown in panel (b), given the
viewing angles derived in Paper~I. {\bf (b; top right)} concentric
ellipses in the $(x'',y'')$ plane of the LMC disk with fixed position
angle and ellipticity chosen to match the observed outer contours
(cf.~Figure~\ref{f:ellparFO}). Panels~(a) and~(b) can be compared to
the bottom two panels of Figure~\ref{f:maps}. {\bf (c; middle left)}
The outermost contour from panel (a), with the major and minor axis of
the ellipse that best describe it. The contour center is shifted with
respect to the center of the inner contours. {\bf (d; middle right)}
The offset $r_{\rm offset}$ of the center of the projected contours as
function of semi-major axis length. {\bf (e; bottom left)} Schematic
representation of various directions on the $(X,Y)$ plane of the sky:
the line of nodes and the perpendicular line that marks the directions
towards the near and the far side of the LMC plane (heavy solid
lines); the direction in which the contour center drifts as a result
of viewing perspective (dash-dotted line); the projection of the
intrinsic major axis of the LMC disk (long-dashed line); and the major
axis of the projected contours on the sky (dotted
line).\label{f:perspec}}

\newcommand{\figcapschemview}
{Schematic representation of the Magellanic system on the projected
$(X,Y)$ plane of the sky, using the zenithal equidistant projection
defined in Section~\ref{ss:coords}. Dotted curves indicate contours of
constant right ascension and declination, as labeled. The LMC is
represented as an ellipse with the size, ellipticity and position
angle of the outermost contour in the near-IR map shown in
Figure~\ref{f:maps}c. The SMC and Galactic Center are indicated as
dots. The Magellanic Stream and the Leading Arm are shown as dashed
curves, the approximate positions of which were traced by eye from the
HI maps of Br\"uns, Kerp \& Staveley-Smith (2000) and Putman \etal
(1998). The lines that start at the centers of the LMC and the SMC
indicate the directions of the proper motions of these systems; the
line-lengths correspond to motion in a period of $5 \times 10^7$
years. The proper motions are averages from a variety of studies as
quoted in Kroupa \& Bastian (1997), and are corrected for the reflex
of the solar motion with respect to the local standard of rest, and
for the motion of the local standard of rest around the Galactic
Center.\label{f:schemview}}

\newcommand{\figcapschemviewthreeD}
{Schematic representation of the Magellanic system in the
three-dimensional $(x'',y'',z'')$ coordinate system defined in
Section~\ref{ss:coordfaceon}. {\bf (a; left)} top view of the LMC disk
plane. The LMC is represented as an ellipse with the size, ellipticity
and position angle of the outermost contour in the deprojected near-IR
map shown in Figure~\ref{f:maps}d. The SMC and Galactic Center are
indicated as dots. The lines that start at the centers of the LMC and
the SMC indicate the directions of their velocity vectors; the
line-lengths correspond to motion in a period of $5 \times 10^7$
years. The velocity vectors are obtained through combination of the
observed radial velocities and proper motions (Kroupa \& Bastian
1997), and are corrected for the reflex of the solar motion with
respect to the local standard of rest, and for the motion of the local
standard of rest around the Galactic Center. {\bf (b; right)} side
view of the LMC disk plane. The positions of the Magellanic Stream and
the Leading Arm are not accurately known in the $(x'',y'',z'')$
coordinate system, because their distances are not well
known.\label{f:schemviewthreeD}}


\ifsubmode
\figcaption{\figcapcmds}
\figcaption{\figcapmaps}
\figcaption{\figcapellpar}
\figcaption{\figcapellparFO}
\figcaption{\figcapperspec}
\figcaption{\figcapschemview}
\figcaption{\figcapschemviewthreeD}

\clearpage
\else\printfigtrue\fi

\ifprintfig


\clearpage
\begin{figure}
\epsfxsize=0.6\hsize
\centerline{\epsfbox{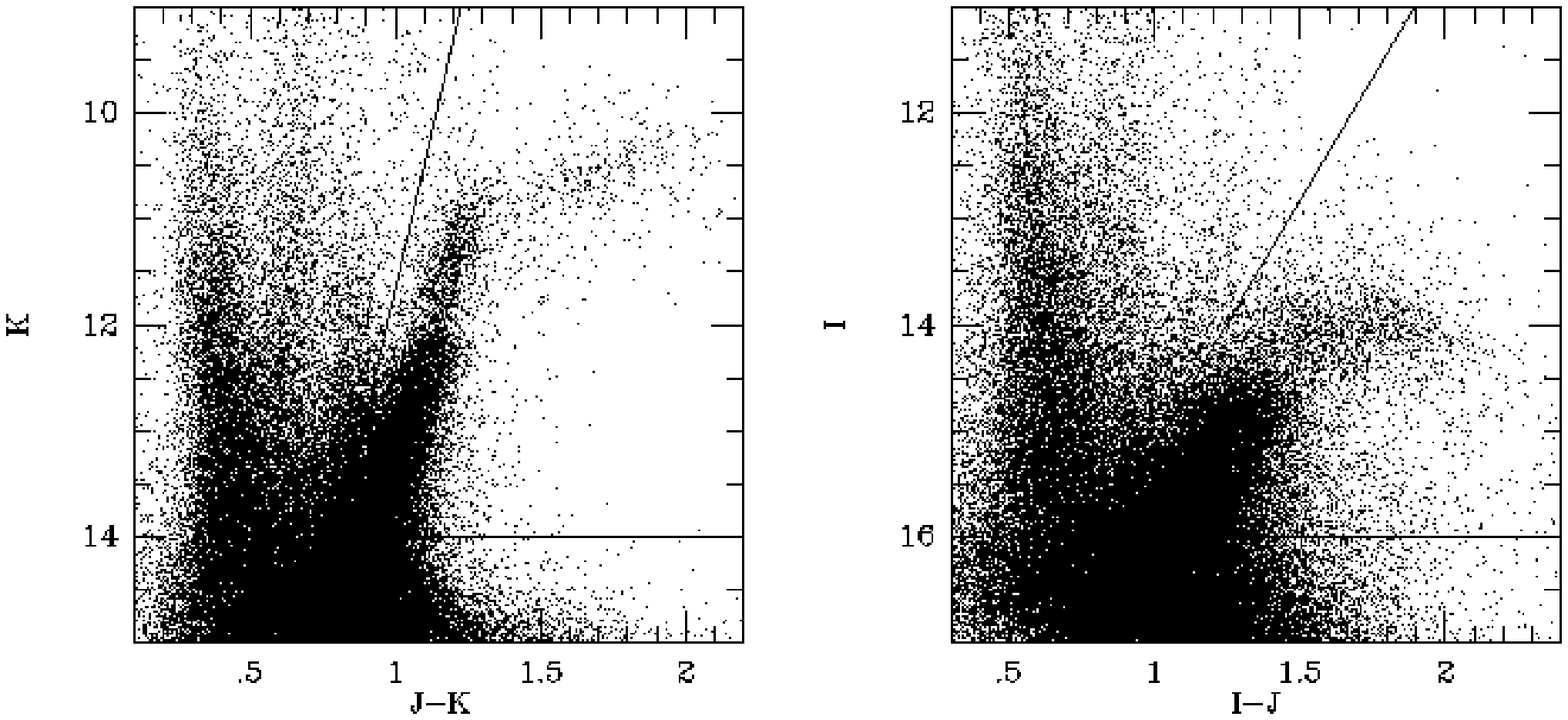}}
\ifsubmode
\vskip3.0truecm
\setcounter{figure}{0}
\addtocounter{figure}{1}
\centerline{Figure~\thefigure}
\else
\figcaption{\figcapcmds}
\fi
\end{figure}


\clearpage
\begin{figure}
\centerline{%
\epsfxsize=0.45\hsize
\epsfbox{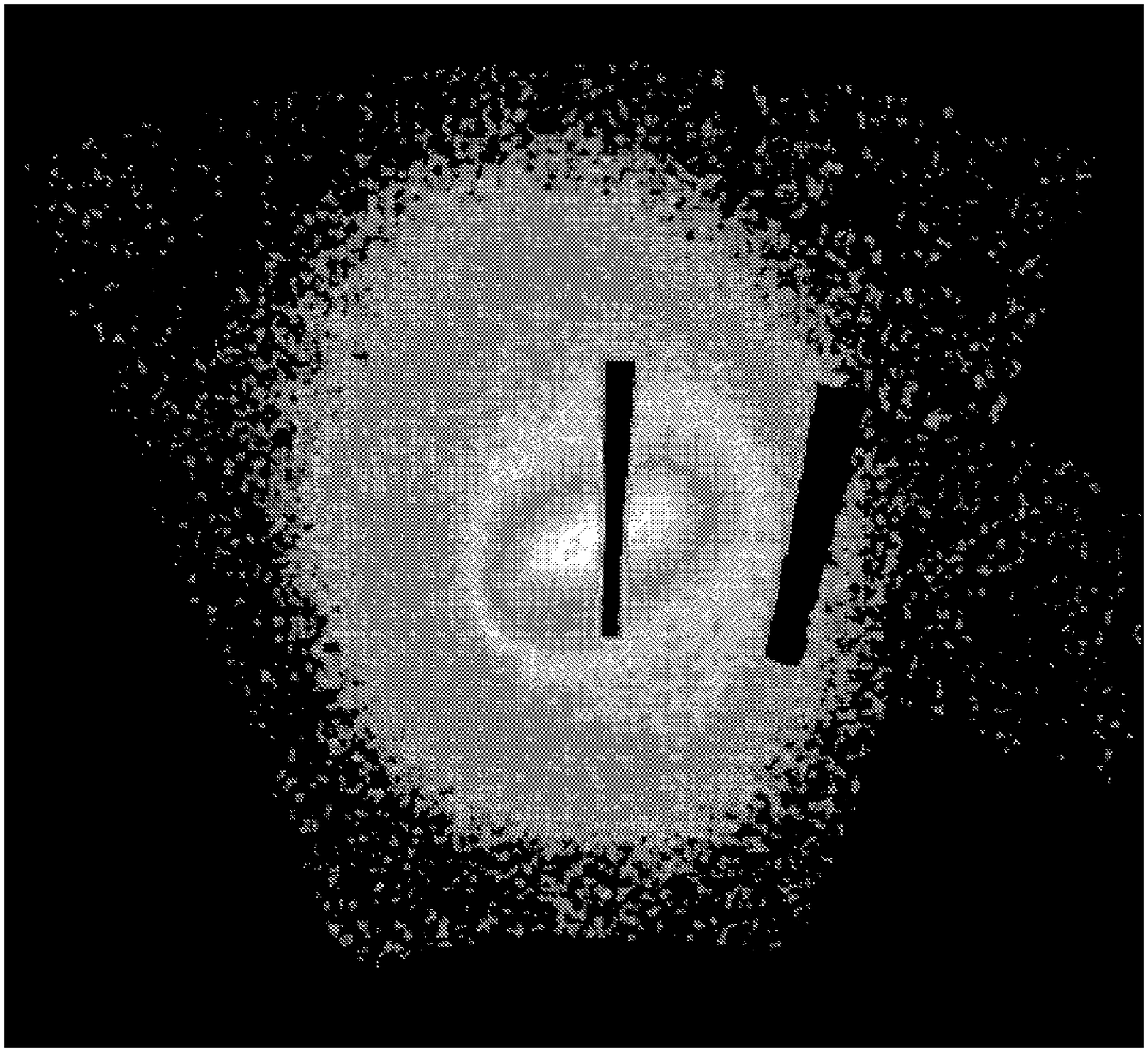}\quad
\epsfxsize=0.45\hsize
\epsfbox{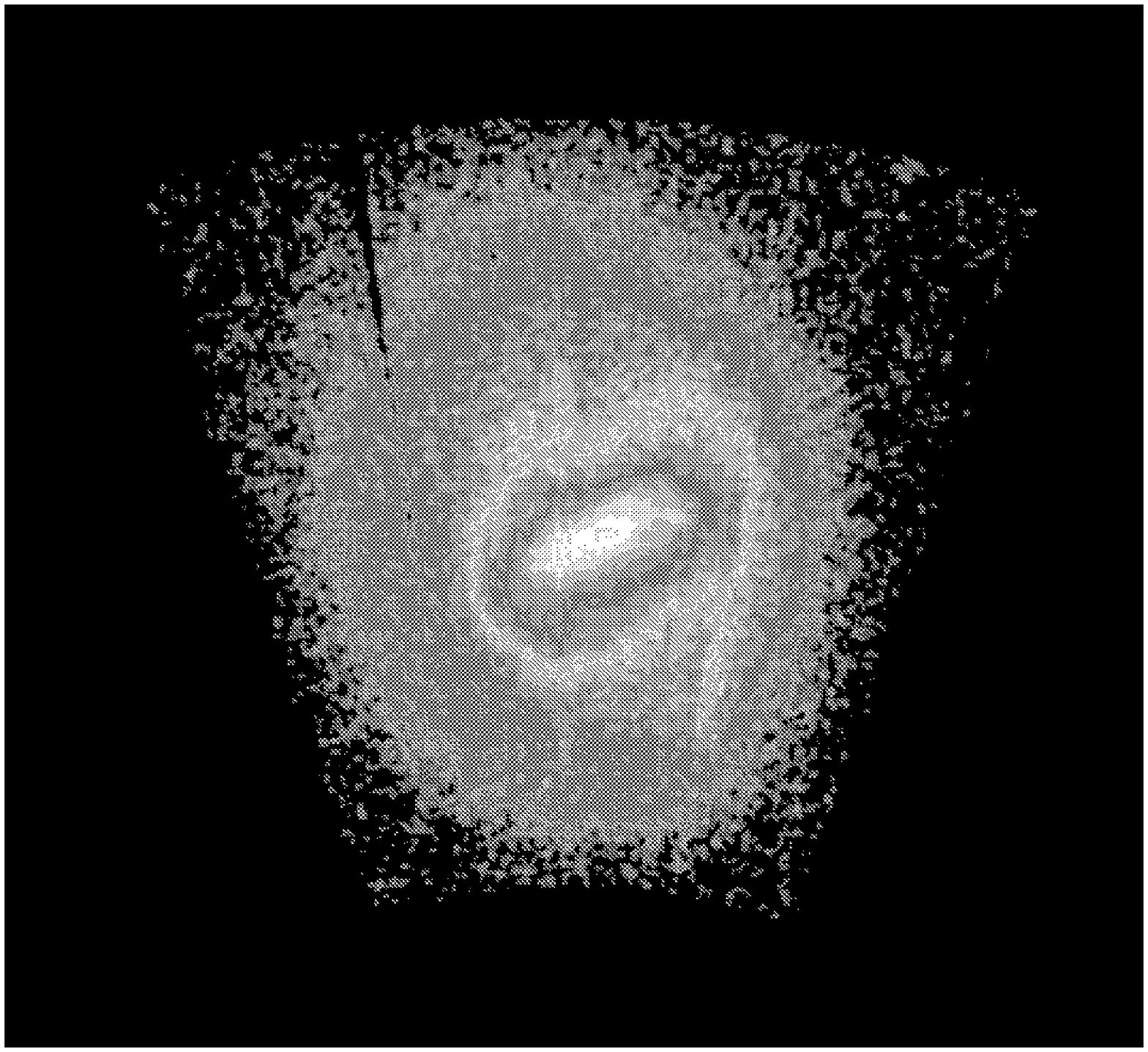}
}
\smallskip
\centerline{%
\epsfxsize=0.45\hsize
\epsfbox{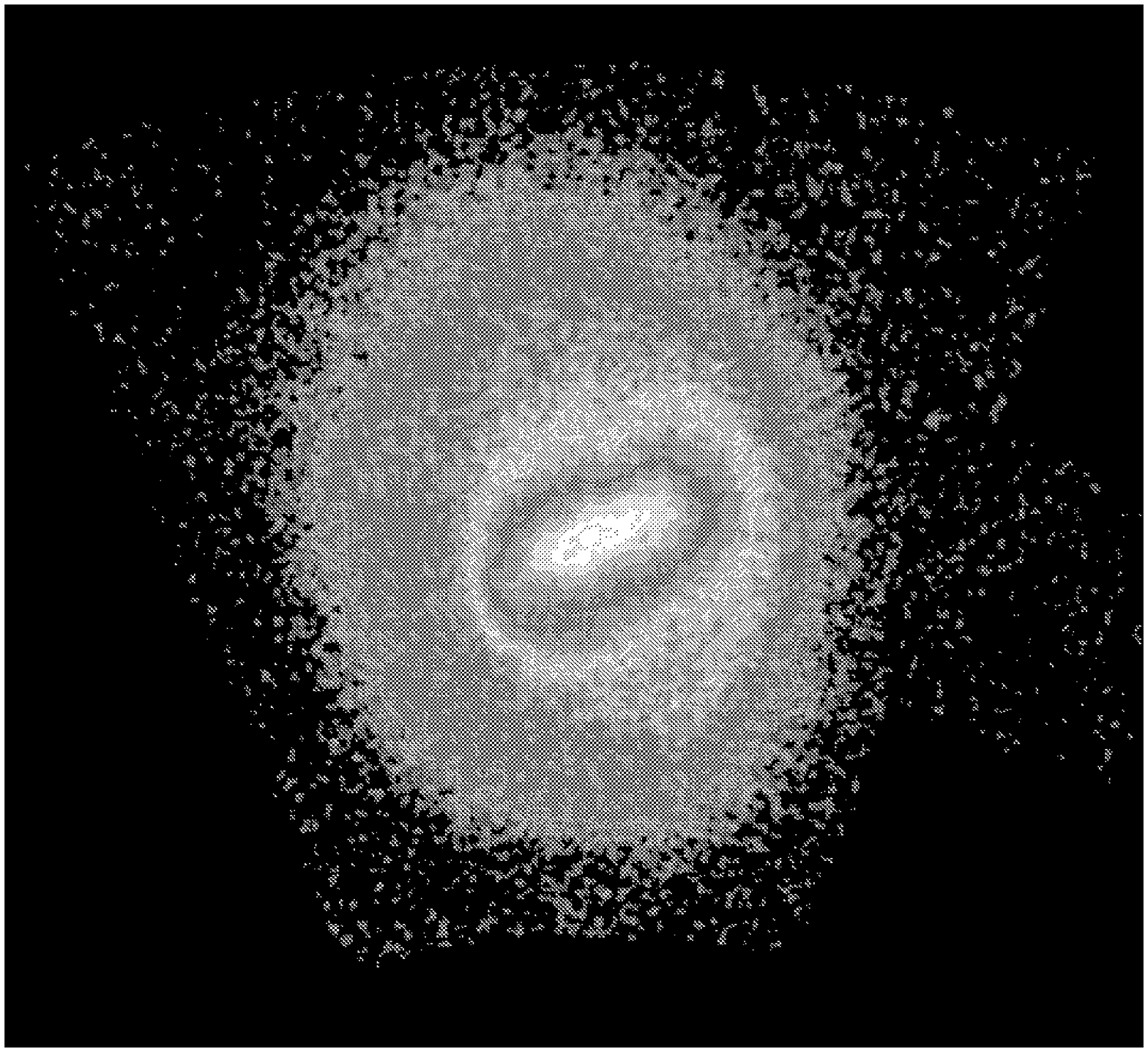}\quad
\epsfxsize=0.45\hsize
\epsfbox{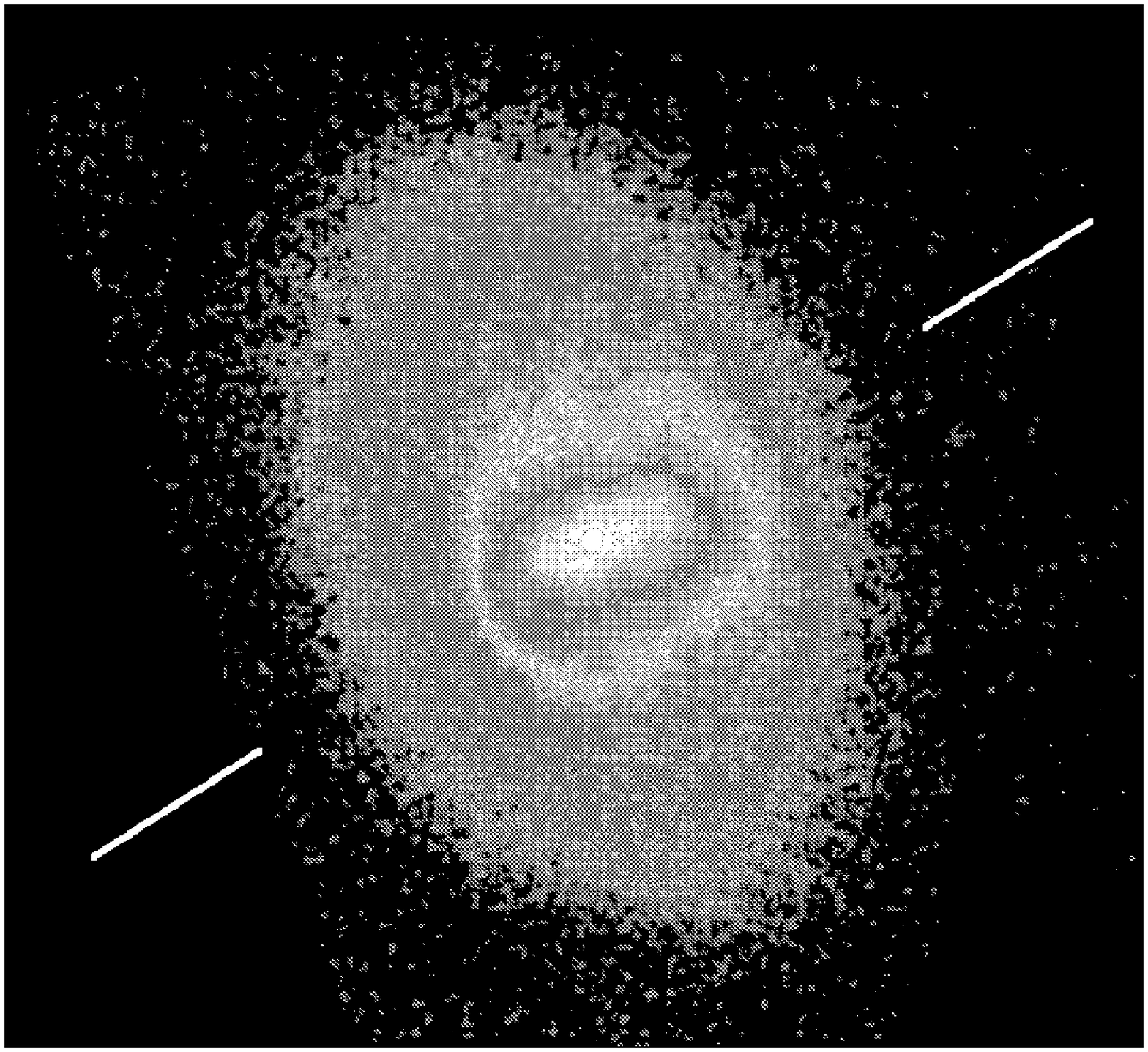}
}
\ifsubmode
\vskip3.0truecm
\addtocounter{figure}{1}
\centerline{Figure~\thefigure}
\else
\figcaption{\figcapmaps}
\fi
\end{figure}


\clearpage
\begin{figure}
\epsfxsize=0.62\hsize
\centerline{\epsfbox{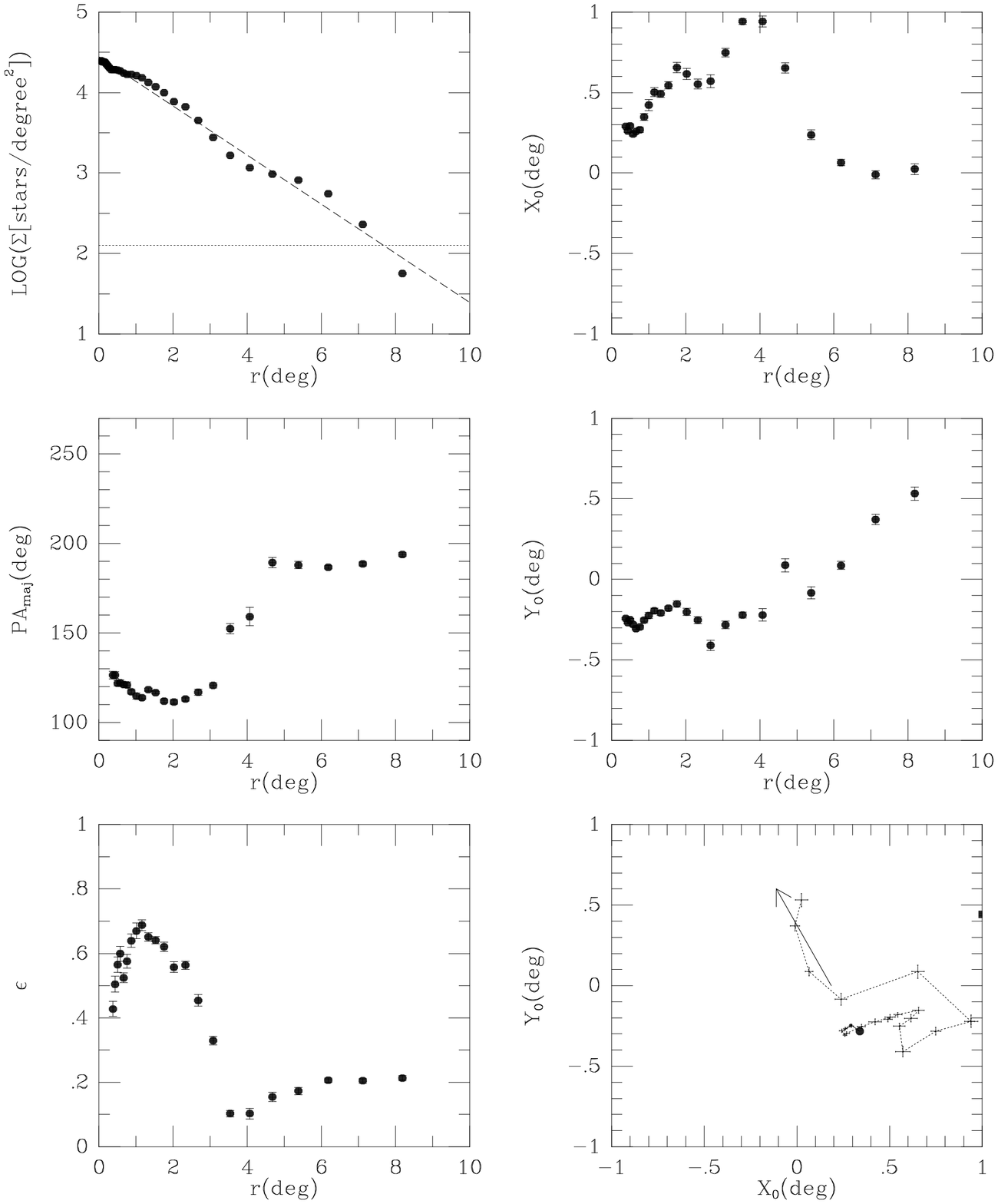}}
\ifsubmode
\vskip3.0truecm
\addtocounter{figure}{1}
\centerline{Figure~\thefigure}
\else
\figcaption{\figcapellpar}
\fi
\end{figure}


\clearpage
\begin{figure}
\epsfxsize=0.62\hsize
\centerline{\epsfbox{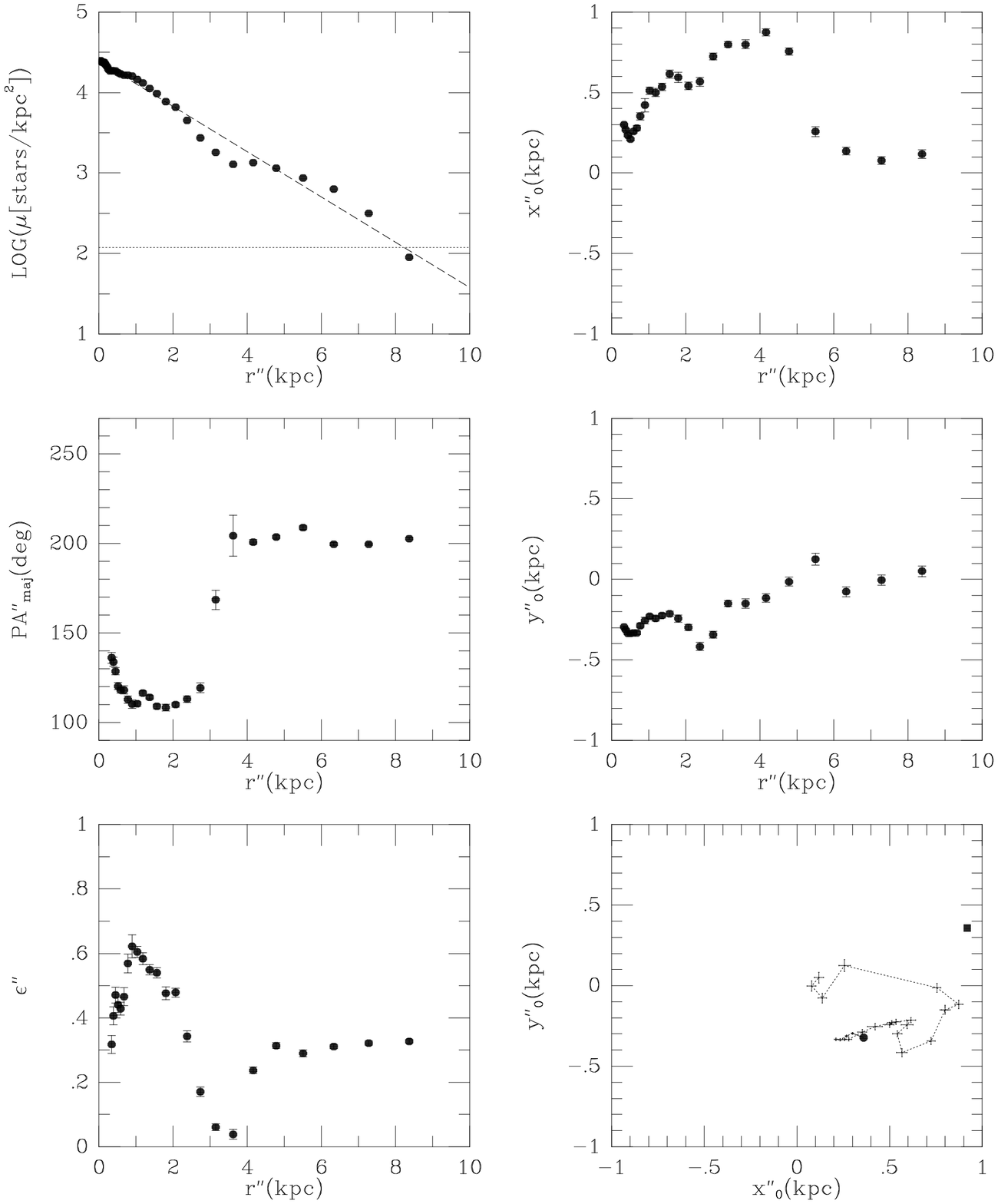}}
\ifsubmode
\vskip3.0truecm
\addtocounter{figure}{1}
\centerline{Figure~\thefigure}
\else
\figcaption{\figcapellparFO}
\fi
\end{figure}


\clearpage
\begin{figure}
\epsfxsize=0.635\hsize
\centerline{\epsfbox{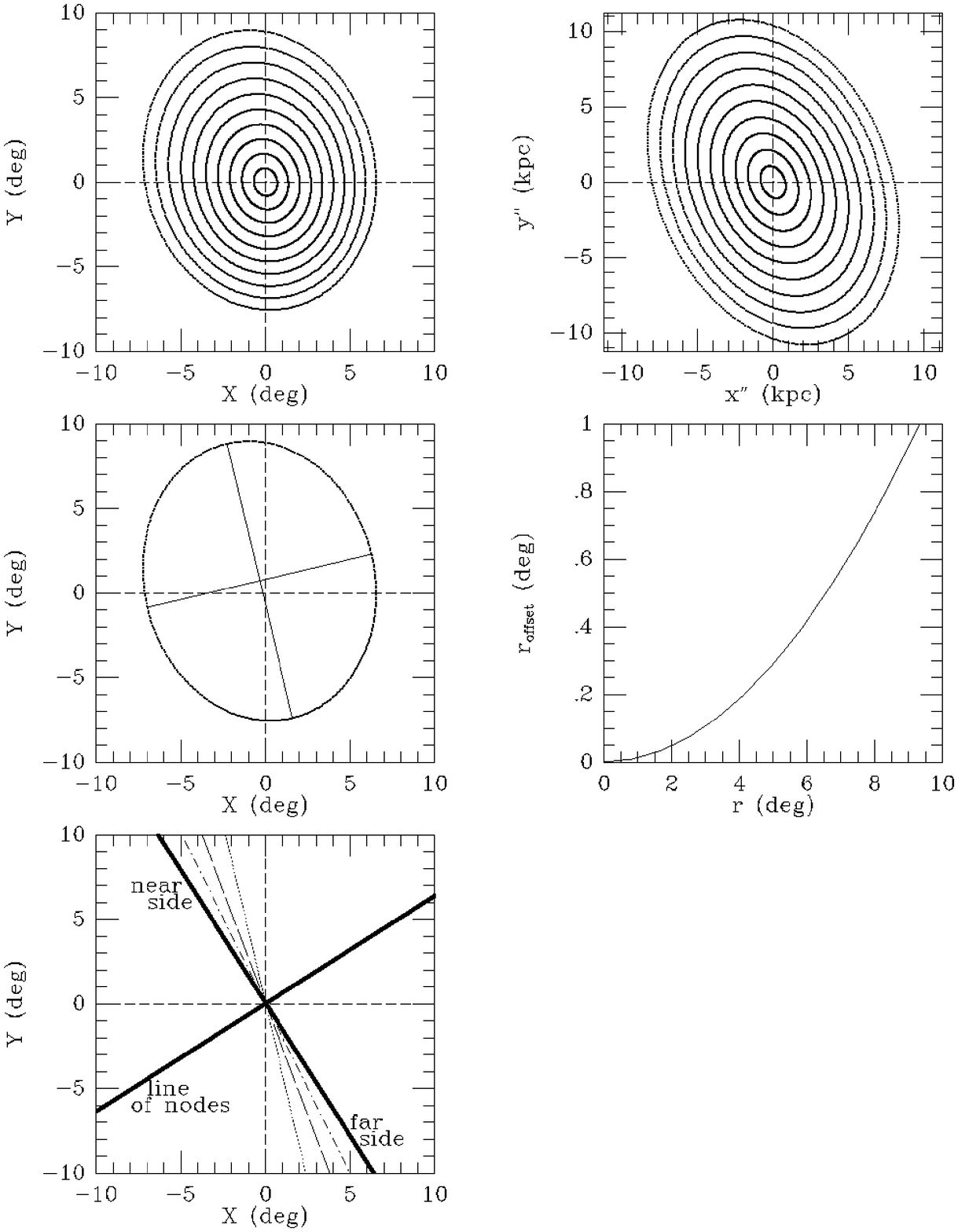}}
\ifsubmode
\vskip3.0truecm
\addtocounter{figure}{1}
\centerline{Figure~\thefigure}
\else
\figcaption{\figcapperspec}
\fi
\end{figure}


\clearpage
\begin{figure}
\epsfxsize=0.9\hsize
\centerline{\epsfbox{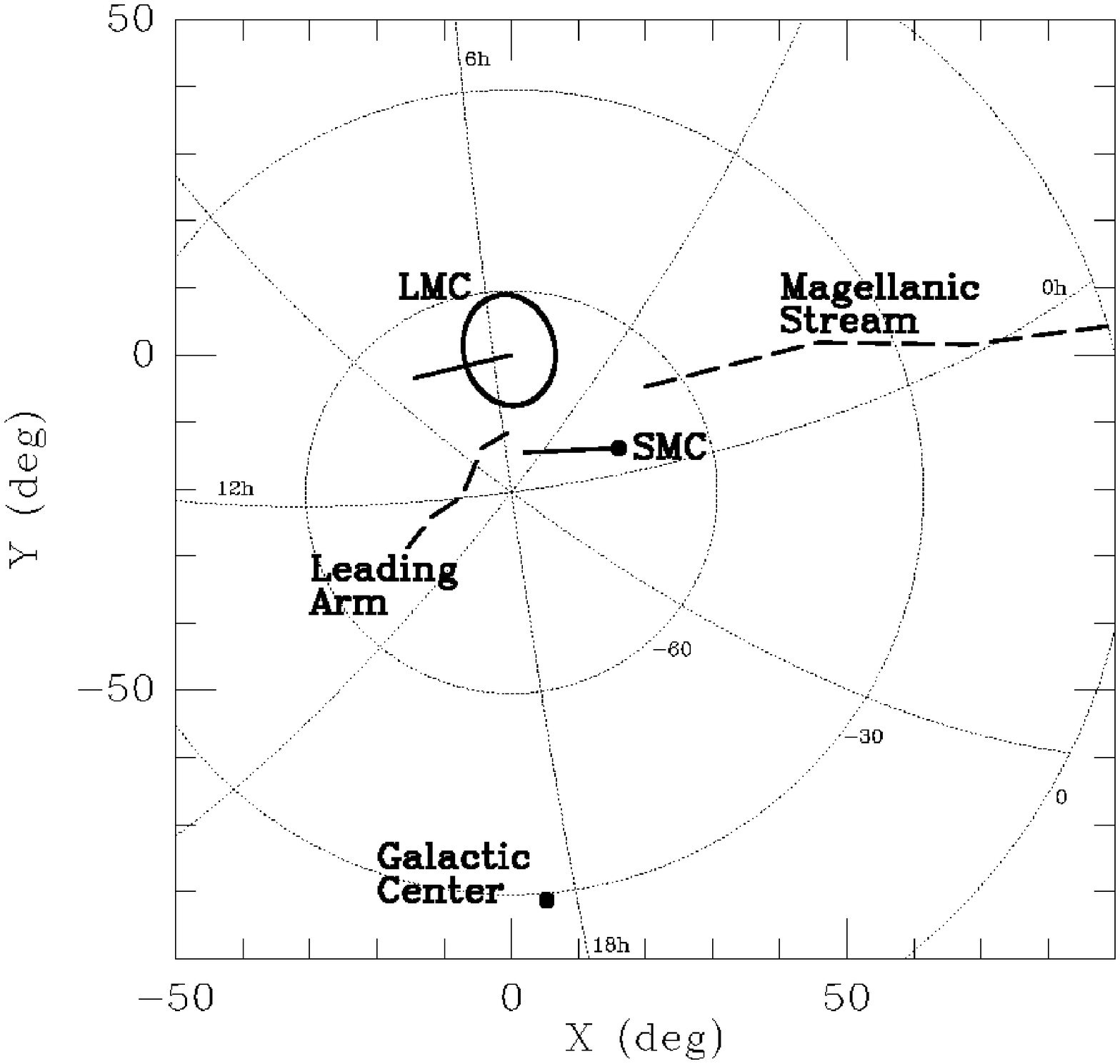}}
\ifsubmode
\vskip3.0truecm
\addtocounter{figure}{1}
\centerline{Figure~\thefigure}
\else
\figcaption{\figcapschemview}
\fi
\end{figure}


\clearpage
\begin{figure}
\epsfxsize=0.9\hsize
\centerline{\epsfbox{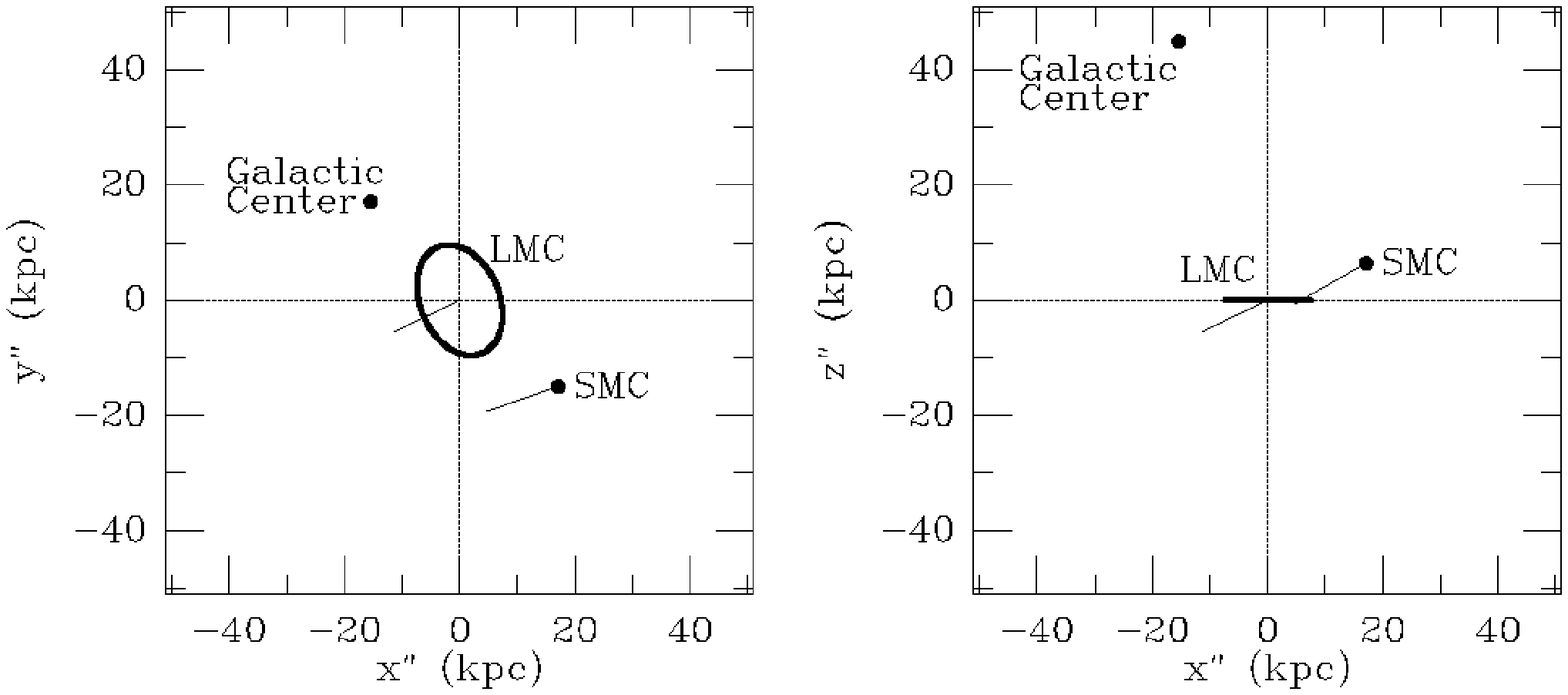}}
\ifsubmode
\vskip3.0truecm
\addtocounter{figure}{1}
\centerline{Figure~\thefigure}
\else
\figcaption{\figcapschemviewthreeD}
\fi
\end{figure}


\fi


\clearpage
\ifsubmode\pagestyle{empty}\fi




\end{document}